\documentclass[1p]{elsarticle}
\usepackage{amsmath}
\usepackage{amsfonts}
\usepackage{amssymb}
\usepackage{xcolor}
\usepackage{hyperref}

\newcommand{\tmu}{\tilde{\mu}_x}
\newcommand{\tDL}{\tilde{D}_\mathrm{max}}
\newcommand{\tsigma}{\tilde{\sigma}^2_x}
\newcommand{\tC}{\tilde{\mathcal{C}}}

\begin{document}

\begin{frontmatter}

\title{A Laplacian Approach to Stubborn Agents and their Role in Opinion Formation on Influence Networks}

\author{Fabian Baumann \textsuperscript{1}, Igor M. Sokolov \textsuperscript{1,2} and Melvyn Tyloo \textsuperscript{3,4}}
\address{\textsuperscript{1} Institut f\"ur Physik and
\textsuperscript{2} IRIS Adlershof, Humboldt-Universit\"at zu Berlin,\\ 
Newtonstra\ss e 15, 12481 Berlin, Germany.}

\address{\textsuperscript{3} Institute of Physics, \'Ecole Polytechnique F\'ed\'erale de Lausanne (EPFL), \\ CH-1015 Lausanne, Switzerland and\\ \textsuperscript{4} School of Engineering, University of Applied Sciences of Western Switzerland HES-SO\\ CH-1951 Sion, Switzerland}

\begin{abstract}
Within the framework of a simple model for social influence, the Taylor model, we analytically investigate the role of stubborn agents in the overall opinion dynamics of networked systems. Similar to zealots, stubborn agents are biased towards a certain opinion and have a major effect on the collective opinion formation process. Based on a modified version of the network Laplacian we derive quantities capturing the transient dynamics of the system and the emerging stationary opinion states. In the case of a single stubborn agent we characterize his/her ability to coherently change a prevailing consensus. For two antagonistic stubborn agents we investigate the opinion heterogeneity of the emerging non-consensus states and describe their statistical properties using a graph metric similar to the resistance distance in electrical networks. Applying the model to synthetic and empirical networks we find while opinion diversity is decreased by small-worldness and favored in the case of a pronounced community structure the opposite is true for the coherence of opinions during a consensus change.
\end{abstract}


\end{frontmatter}


\section{Introduction}
The basic assumption of social influence theory suggests that people adapt their opinions, attitudes or conventions collectively upon interactions  \cite{kelman1958compliance}. 
Hence, uncovering the dynamics towards global opinion states yields an important cornerstone in the understanding of social dynamics that has been addressed by a multitude of studies \cite{baronchelli2018emergence, castellano2009statistical}. 
Recently it has been shown empirically that social conventions might be turned over by small committed minorities in a population \cite{centola2018experimental}. Moreover, many theoretical studies showed that the presence of stubborn agents or zealots, which do not change their opinions, can crucially change the collective dynamics of such systems. In particular, those studies considered the voter model \cite{mobilia2003does,mobilia2007role,yildiz2011discrete,klamser2017zealotry,yildiz2013binary,masuda2015opinion} as a simple approach to discrete opinion dynamics and the naming game \cite{verma2014impact,xie2011social,waagen2015effect} to model the dynamics of more general social conventions.

In the field of opinion dynamics a special focus traditionally lies on the formation of consensus \cite{abelson1964mathematical,deffuant2000mixing,hegselmann2002opinion,baronchelli2018emergence} as well as the emergence of heterogeneous (non-consensus) states \cite{banisch2019opinion,dandekar2013biased,kurahashi2016robust}. This is due to the supposed relevance of such opinion states on a societal level \cite{castellano2009statistical,baronchelli2018emergence}. 
Remarkably, it has been shown that even in the case of infinitely large systems single zealots have finite effects on the overall dynamics \cite{mobilia2003does} and are, more specifically, able to preclude a global consensus in the system \cite{mobilia2007role}. Also more specific questions were addressed. For instance in Ref.~\cite{yildiz2011discrete} it was studied which positions in the network maximize the influence of stubborn agents on the total population. Similar results have been obtained for the naming game. Here, a small committed minority of zealots was able to rapidly reverse a prevailing majority \cite{xie2011social} 
or to restrict certain opinions to a small subset of nodes \cite{waagen2015effect}. 

In contrast to Refs.~\cite{mobilia2003does,mobilia2007role,yildiz2011discrete,klamser2017zealotry,yildiz2013binary,masuda2015opinion,verma2014impact,xie2011social,waagen2015effect,arendt2015opinions}, we consider stubborn agents in the context of continuous opinion formation. We aim to quantify their influence on given networks and based on their positions therein. More specifically, we consider the following two paradigmatic situations:
$(i)$  the dynamics of consensus change induced by a single stubborn agent, and $(ii)$ the emergence of non-consensus states due to a pair of antagonistically biased agents.
In the first case we focus on how the consensus change happens and aim to quantify the persuasiveness of single stubborn agents. Inspired by recent studies in network control theory \cite{patterson2010leader,mateo2019optimal}, we derive a measure for network opinion coherence. It predicts, based on the positions of stubborn agents, how closely the population will follow their opinions during the transient dynamics.
In the second case we investigate the characteristic properties of stationary opinion states. As previously shown for the voter model \cite{yildiz2011discrete}, the emerging states deviate strongly from a full consensus and their specific properties crucially depend on the positions of the two antagonistic stubborn agents.

To model the collective dynamics of continuous opinion exchange in the presence of stubborn agents, we use the Taylor model \cite{taylor1968towards}. Based on the seminal work of Abelson \cite{abelson1964mathematical} it is similar to various previous approaches of constructive opinion dynamics \cite{deffuant2000mixing,hegselmann2002opinion} and represents a minimal model for social interactions. It proposes a diffusive coupling between agents' opinions. 
Crucially, some individuals referred to as stubborn agents, 
are additionally influenced by individual biases, which are usually interpreted as strong personal prejudices or cues from external communication sources \cite{proskurnikov2017tutorial,taylor1968towards}.
Although the linear dynamics of the Taylor model appears to be rather simplistic, its discrete time version, introduced by Friedkin and Johnsen \cite{friedkin1990social}, has previously been validated in experimental studies for small and medium sized groups \cite{proskurnikov2017tutorial,childress2012cultural,friedkin2011social}.
Hence, our theoretical considerations shed light on a simple mechanism of opinion exchange, which might underlie actual social phenomena. Those include the targeted disruption of an established societal consensus, e.g. on climate change \cite{lewandowsky2019science}, or an increased opinion heterogeneity promoting social polarization around controversial topics \cite{fiorina2008political,bail2018exposure}. 

Based on a modified version of the network Laplacian, we investigate the role of stubborn agents using a spectral decomposition. It allows us to express properties of the transient dynamics  and the heterogeneity of the emerging non-consensus states in terms of compact closed expressions that depend on the positions of stubborn agents. Crucially, those quantities can be formulated in terms of a novel graph-based metric, similar to the resistance distance \cite{Kle93}. This formulation allows to probe the interplay between the position of stubborn agents and properties of the emerging non-consensus states using intuitive arguments, that are related to the structure of the influence network. We demonstrate this on   Watts-Strogatz (WS) \cite{watts1998collective} and stochastic block model (SBM) \cite{holland1983stochastic} networks to focus on network features which have previously been shown to affect the properties of dynamical processes \cite{watts1998collective,gargiulo2010opinion,mistry2015committed}.

The paper is organized as follows. In Sec.~\ref{sec:leader-follower-dynamics} the dynamical equations of the Taylor model are introduced. Subsequently, in Sec.~\ref{sec3} we discuss important aspects of the Laplacian formalism and introduce the concept of modified resistance distances. In Sec.~\ref{subsec:opinion-coherence} and  Sec.~\ref{sec:two-opposing-leaders} we analytically discuss paradigmatic cases of stubborn agents on influence networks and numerically evaluate those results in Sec.~\ref{sec:num} on different network topologies. The work is concluded in Sec.~\ref{sec:conclusion}.

\section{The Model}\label{sec:leader-follower-dynamics}
The Taylor model considers a system of $n$ interacting agents. Each agent $i$ is characterized by a continuous opinion variable $x_i\in\mathbb{R}$. The dynamics of opinion exchange are governed by the following set of differential equations,
\begin{subequations}\label{eq:LF}
\begin{align}\label{eq:leaders_basic}
\dot{x}_i&=-\sum_j b_{ij}(x_i-x_j) \; , \; \; i\not\in V_s \;,\\
\label{eq:followers_basic}
\dot{x}_i&=-\sum_j b_{ij}(x_i-x_j) -\kappa [x_i-P_i(t)] \; , \; \; i\in V_s \; ,
\end{align}
\end{subequations}
where $b_{ij}$ denotes the elements of the adjacency matrix $\mathbf{B}$ encoding the influence network consisting of $n$ nodes and $n_e$ edges. The set of stubborn agents is given by $V_s$. We assume symmetric and non-repulsive interactions, such that $b_{ij}=b_{ji}$ and $b_{ij}\geq0$ if agents $i$ and $j$ are interacting, and $b_{ij}=0$ otherwise. Due to the first terms in Eqs.~(\ref{eq:LF}), all agents aim to minimize the opinion differences to their connected neighbors. Crucially, the opinion of a stubborn agent $a$, $x_a$, is additionally influenced towards his/her bias $P_a$, modeled by the last term of Eq.~(\ref{eq:followers_basic}). The parameter $\kappa$ controls the rate of convergence towards the bias and is therefore termed stubbornness. Note that for $\kappa\rightarrow0$ stubborn agents do not follow their biases and behave as regular agents. Extending previous research efforts \cite{Pat18}, we assume that bias values $P_i$ are not constant across stubborn agents.

\begin{figure}[h]
\includegraphics[width=0.99\linewidth]{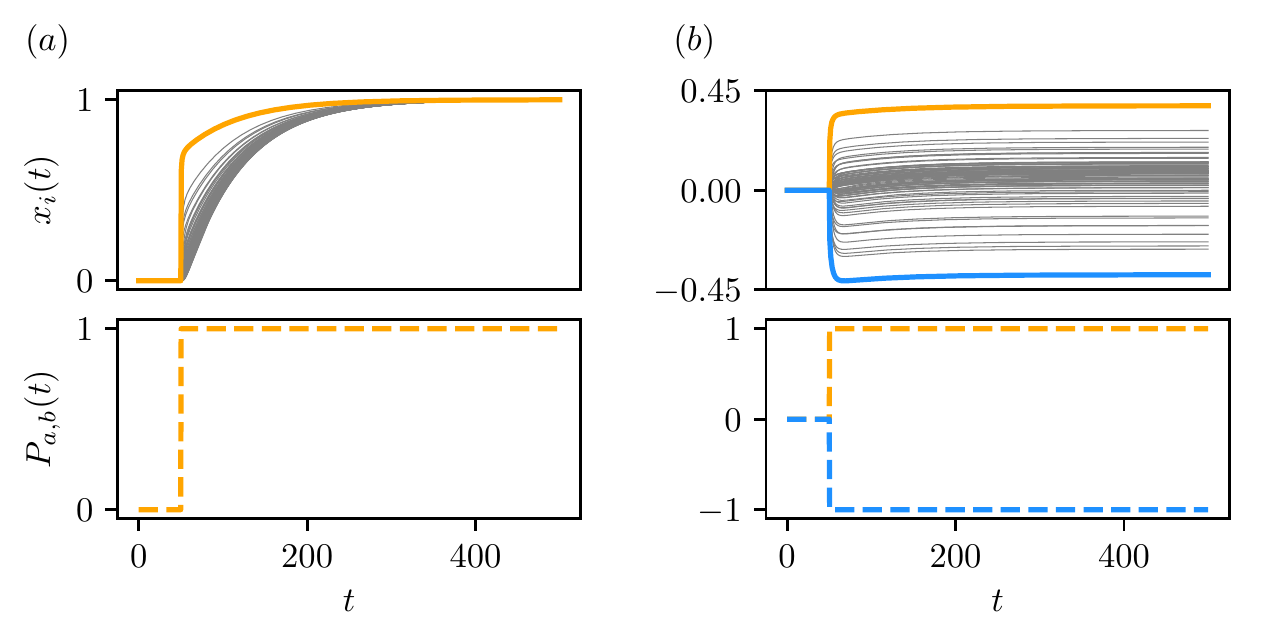}
\caption{Opinion dynamics for different sets of stubborn agents. In both depicted cases the system is initialized at a perfect consensus at $P_i(0)=x_i(0)=0\,$ $\forall i$. At $t_0=50$ the stubborn agents are assumed to develop biases towards a new value $P_i\neq0$. The opinions are shown in the top panels and the biases of stubborn agents are depicted in the bottom panels.
In the case of a single stubborn agent $i=a$ [panel (a)], the system changes its initial consensus at $x_i=0$ to a new opinion value specified by $P_a(t)$. The opinion of the stubborn agent $x_a$ (thick orange dashed line) and the opinions of regular agents (gray thin solid lines) converge to the bias of agent $a$.
For two opposing stubborn agents with $P_a(t\geq t_0) = 1 = - P_b(t\geq t_0)$  the system does not reach a new consensus [panel (b)]. Instead, the opinions of the remaining agents (thin gray solid lines) are distributed between the final opinions of the positive (thick orange dashed line) and the negative (thick blue solid line) stubborn agent. Note that for $n_s>1$ [i.e. panel (b)] the opinions of stubborn agents do generally not reach their individual biases, i.e. $|x_i(t\rightarrow\infty)|\leq |P_i(t\geq t_0)|$.  
The underlying influence network $b_{ij}$ is modeled as an unweighted Watts-Strogatz (WS) graph with $n=100$, $K_\mathrm{WS}=4$ and $p_r=0.1$. The stubbornness parameter was set to $\kappa=1$.}
\label{fig:fig1}
\end{figure}

Figure~\ref{fig:fig1} illustrates the two specific situations, which we will consider in this work, i.e. the case of a single stubborn agent $V_s=\{a\}$ [Fig.~\ref{fig:fig1}(a)] and a pair of antagonistically biased stubborn agents $V_s=\{a,b\}$ [Fig.~\ref{fig:fig1}(b)]. In both cases, starting at $x_i(0)=P_i(0)=0\,\,\forall i$, we assume that, at $t=t_0$, the stubborn agents develop biases which are not aligned with the prevailing consensus in the system, i.e. $P_i(t\geq t_0)\neq0$.

For a single stubborn agent $(V_s=\{a\})$, the system is driven towards a new consensus. We observe that, after a transient period, all opinions $x_i$ converge to the new bias of agent $a$. Our analysis will focus on the opinion coherence during the consensus change, which strongly depends on the topology of the influence network and the position of the stubborn agent.

In the case of a pair of antagonistically biased agents ($V_s=\{a,b\}$), with biases assumed as $P_a(t)=-P_b(t)$, the system does not reach a new consensus, cf. Fig.\ref{fig:fig1}(b). Instead, opinions are finally distributed over an interval of finite width. In this case we aim to characterize the heterogeneity of the emerging non-consensus states depending on the placement of biased agents.  

In the following we formulate the model equations in terms a modified version of the network Laplacian. Below we present important properties of the corresponding operator and demonstrate how it is utilized to analytically solve the dynamics of the system using a spectral decomposition. Subsequently, we introduce the concept of modified resistance distances (MRD). Those MRDs will play a crucial role in the formal description of the stationary properties of the system in the cases of antagonistically biased agents.

\section{Laplacian formalism and Modified Resistance Distances}\label{sec3}
Defining the opinion vector ${\bf x}=(x_1,...,x_n)$, where the system size $n$ is the sum of the number of stubborn ($n_s$) and regular agents, Eqs.~(\ref{eq:LF})
can be rewritten in vectorial form as
\begin{eqnarray}\label{eq:leafolvec}
\dot{{\bf x}}=-(\mathbb{L} + K)\,{\bf x} + \kappa{\bf P} \; .
\end{eqnarray}
Here, $K$ is a diagonal matrix with $K_{ii} = \kappa\,\,$ for$\,\, i \in V_s$ and vanishing components for $i\not\in V_s$. Similarly, the time-dependent vector ${\bf P}(t)$ contains the bias opinions of stubborn agents with vanishing elements $P_i(t)=0$ if $i\not\in V_s$. The purely diffusive part of the agents' interactions can be expressed using the regular Laplacian $\mathbb{L}$ of the coupling graph. It is defined as 
\begin{equation}\label{eq:laplacian}
{\mathbb L}_{ij} = 
\left\{ 
\begin{array}{cc}
-b_{ij}  \, , & i \ne j \, , \\
\sum_k b_{ik}  \, , & i=j \,.
\end{array}
\right.
\end{equation}
We consider undirected influence networks. Hence, both $\mathbb{L}$ and $\mathbb{L}^{(\kappa)}=(\mathbb{L}+K)$ are symmetric matrices. In the following we refer to $\mathbb{L}^{(\kappa)}$ as the modified Laplacian. 

The system of Eqs.~\eqref{eq:LF} can be solved by a spectral decomposition of $\bf x$ over the eigenvectors $\mathbf{u}^{(\kappa)}_\alpha$ of $\mathbb{L}^{(\kappa)}$, i.e.
\begin{equation}\label{eq:spec-decomp}
x_i(t)=\sum_{\alpha}c_\alpha(t)\, u^{(\kappa)}_{\alpha,i}\,,
\end{equation}
where $c_\alpha(t)$ and $u^{(\kappa)}_{\alpha,i}$ denote the time-dependent coefficients of the expansion and the $i$-th element of the $\alpha$-th eigenvector of $\mathbb{L}^{(\kappa)}$, respectively. Substituting Eq.~\eqref{eq:spec-decomp} into Eq.~\eqref{eq:leafolvec} gives the following set of differential equations for the expansion coefficients, 
\begin{eqnarray}\label{eq:expansion_coefficients}
\dot{c}_\alpha = -\lambda_\alpha^{(\kappa)} \, c_\alpha + \kappa\,{\bf P}\cdot{\bf u}_\alpha^{(\kappa)} \;,\quad \alpha=1,\ldots, n\;,
\end{eqnarray}
where $\lambda_\alpha^{(\kappa)}$ denotes the $\alpha$-th eigenvalue of 
$\mathbb{L}^{(\kappa)}$.
The general solution to Eq.~\eqref{eq:expansion_coefficients} then reads
\begin{align}\label{eq:sol}
\begin{split}
c_\alpha(t)=c_\alpha(0)e^{-\lambda_\alpha^{(\kappa)} t} +\kappa\, e^{-\lambda_\alpha^{(\kappa)} t}\int_0^t e^{\lambda_\alpha^{(\kappa)} t'}\, {\bf P}\cdot{\bf u}_\alpha^{(\kappa)}\, {\rm d}t'\;  \, \,,
\end{split}
\end{align}
for $\alpha = 1,\dots,n$. Solving Eq.~(\ref{eq:sol}) for a specific vector of bias opinions ${\bf P}$ finally yields the time-evolution ${\bf x}(t)$ of the system.

\subsection{Properties of $\mathbb{L}^{{(\kappa)}}$}\label{secL}
Depending on $n_s$, the modified Laplacian $\mathbb{L}^{(\kappa)}$ and its associated eigenvectors 
have specific properties. In the following we discuss those that are relevant for the cases of one and two stubborn agents. Elementwise $\mathbb{L}^{(\kappa)}$ reads,
\begin{eqnarray}
\mathbb{L}^{(\kappa)}_{ij}=\mathbb{L}_{ij} + \kappa\sum_{a\in V_s}\delta_{ij}\delta_{ia} \; ,
\end{eqnarray}
where 
$\mathbb{L}$ satisfies $\sum_j \mathbb{L}_{kj}=0$ $\forall k$.
The elements of the inverse of $\mathbb{L}^{(\kappa)}$ along rows and columns corresponding to nodes of
stubborn agents satisfy accordingly,
\begin{eqnarray}\label{eq:1/k}
\sum_{a\in V_s}[{\mathbb{L}^{(\kappa)}}]^{-1}_{aj}=\sum_{a\in V_s}\sum_\alpha\frac{u_{\alpha,a}^{(\kappa)}\,u_{\alpha,j}^{(\kappa)} }{\lambda_\alpha^{(\kappa)}}=1/\kappa \; 
\end{eqnarray}
which holds true for $n_s\geq 1$ 
(see derivation in \ref{app1}). 
With $[\mathbb{L}^{(\kappa)}]^{-1}_{ij}$ we denote the 
$ij$-th element of the inverse of $\mathbb{L}^{(\kappa)}$.

Particularly, in the case of two biased agents ($V_s=\{a,b\}$) Eq.~(\ref{eq:1/k}) yields the relation
\begin{eqnarray}\label{eq:prop2}
[{\mathbb{L}^{(\kappa)}}]^{-1}_{aa}&=&[{\mathbb{L}^{(\kappa)}}]^{-1}_{bb} \;,
\end{eqnarray}  
as derived in \ref{app2}. Note that for more than two leaders, 
i.e. $n_s>2$, this relation needs to be generalized.

For arbitrary $n_s$ 
the eigenvectors of ${\bf u}_\alpha^{(\kappa)}$
satisfy 
\begin{eqnarray}\label{eq:sum}
\sum_i u_{\alpha,i}^{(\kappa)} =\kappa\sum_{a\in V_s}\frac{u_{\alpha,a}^{(\kappa)}}{{\lambda_\alpha^{(\kappa)}}} \;,
\end{eqnarray}
cf. derivation in \ref{App4}.
This is combined with Eq.~(\ref{eq:1/k}) to obtain the last necessary relation for the 
following considerations as
\begin{eqnarray}\label{eq:ov}
\sum_{a,b\in V_s}[{\mathbb{L}^{(\kappa)}}]^{-2}_{ab}=\sum_{a,b\in V_s}\sum_{\alpha}\frac{u_{\alpha,a}^{(\kappa)} u_{\alpha,b}^{(\kappa)}}{{\lambda_\alpha^{(\kappa)}}^2}=n/\kappa^2 \; .
\end{eqnarray}
The presented properties of $\mathbb{L}^{(\kappa)}$ will be used in the following to derive compact expressions capturing the dynamics of the Taylor model in the two considered settings of $V_s=\{a\}$ and $V_s=\{a,b\}$, respectively. For the latter case, Eqs.~\eqref{eq:1/k}-\eqref{eq:ov}, will be utilized to quantify the properties of heterogeneous opinion states in terms modified resistance distances, which we introduce below.

\subsection{Modified Resistance Distances (MRDs)}\label{sec:resistance_distance}
The resistance distance is a graph-theoretic metric originally based on the Laplacian of an interaction network~\cite{Kle93},
\begin{eqnarray}\label{eq:original-resistance-distance}
\Omega_{ij}=\mathbb{L}_{ii}^\dagger + \mathbb{L}_{jj}^\dagger - \mathbb{L}_{ij}^\dagger - \mathbb{L}_{ji}^\dagger \; ,
\end{eqnarray}
where $\mathbb{L}_{ij}^\dagger$ denotes $ij$-th element of the pseudoinverse of $\mathbb{L}$ defined as
\begin{eqnarray}
\mathbb{L}^\dagger= [\mathbb{L}+\mathbb{I}]^{-1}-n^{-1}\mathbb{I}\; .
\end{eqnarray}
Here, $\mathbb{I}$ represents the $n\times n$ matrix full of ones, i.e. $\mathbb{I}_{ij}=1$ $\forall i,j$\,.
In contrast to $\mathbb{L}$, the modified Laplacian $\mathbb{L}^{(\kappa)}$ is nonsingular (see \ref{app0}).
A distance metric 
similar to $\Omega_{ij}$ can thus be defined 
using the inverse of $\mathbb{L}^{(\kappa)}$, i.e.
\begin{eqnarray}\label{eq:RLd}
\Omega_{ij}^{(\kappa,1)}(V_s)=[{\mathbb{L}^{(\kappa)}}]^{-1}_{ii} + [{\mathbb{L}^{(\kappa)}}]^{-1}_{jj} - [{\mathbb{L}^{(\kappa)}}]^{-1}_{ij} - [{\mathbb{L}^{(\kappa)}}]^{-1}_{ji} \;,
\end{eqnarray}
where the index $1$ denotes the first order modified resistance distance (MRD). It depends on the influence network $\mathbf{B}$, the specific set of stubborn agents $V_s$ and the stubbornness $\kappa$. 
Following \cite{Kle93,Xia03,Tyl18a,Tyl18b,Tyl19}, generalizations of $\Omega_{ij}^{(\kappa,1)}$ to the $p$-th power of the modified Laplacian can be expressed using eigenvectors and eigenvalues of 
$\mathbb{L}^{(\kappa)}$ as, 
\begin{subequations}\label{eq:eig}
\begin{align}
\Omega_{ij}^{(\kappa,p)}(V_s)&=
[{\mathbb{L}^{(\kappa)}}]^{-p}_{ii} + [{\mathbb{L}^{(\kappa)}}]^{-p}_{jj} - [{\mathbb{L}^{(\kappa)}}]^{-p}_{ij} - [{\mathbb{L}^{(\kappa)}}]^{-p}_{ji}\,,\\\label{eq:eig2}
&=\sum_\alpha\frac{(u_{\alpha,i}^{(\kappa)}
-u_{\alpha,j}^{(\kappa)})^2}{{\lambda_\alpha^{(\kappa)}}^p} \; .
\end{align}
\end{subequations}
Using this general-order MRD between nodes $i$ and $j$, we introduce the associated closeness centrality,
\begin{eqnarray}\label{eq:centr}
C_p(i,V_s)=\left[ n^{-1}\sum_j 
\Omega_{ij}^{(\kappa,p)}(V_s) \right]^{-1} \; ,
\end{eqnarray}
quantifying the average MRD of order $p$ from node $i$ to any other node in the influence network \cite{sabidussi1966centrality,Bol14,Tyl18b,Tyl19} given $V_s$. More precisely, if $C_p(i,V_s)$ is large (small), then the agent on node $i$ is central (peripheral) according to the distance defined in Eq.~(\ref{eq:eig}).
Note that if Eq.~(\ref{eq:centr}) is written using Eq.~\eqref{eq:eig2}, properties of the eigenvectors ${\bf u}_\alpha^{(\kappa)}$ which depend on the number of stubborn agents ($n_s$) need to be taken into account.
Importantly, the introduced MRDs and all quantities derived therefrom depend on the specific set of stubborn agents. 
Hence, those quantities need to be recomputed, whenever $V_s$ changes.

\section{Consensus change}\label{subsec:opinion-coherence}

In this section we analytically investigate the first of two prototypical situations -- the dynamics of the system towards a new consensus value. Specifically,  we assume a single stubborn agent driving the population of agents towards his/her bias value. We aim to quantify how closely, or \emph{coherently}, the remaining agents follow the stubborn agent's opinion trajectory. The approach is inspired by previous studies on network coherence in leader-follower systems \cite{patterson2010leader,mateo2019optimal} in which the effects of single leading agents on the collective dynamics are considered. In line with   previous studies, we find that the opinion coherence of the system critically depends on the position of the stubborn agent \cite{Pat18,mateo2019optimal}. In the context of social dynamics the concept of opinion coherence might be utilized to change a prevailing consensus in a desired manner, e.g. to select potentially influential individuals which are closely followed by the population.

We consider the following setup. Initially, the bias $P_a$ of the stubborn agent $a$ is assumed to be compliant with the global consensus, i.e. $P_a(0)=x_i(0)\,\forall\,i$.  Subsequently, however, at $t=t_0$, its value changes towards a specific opinion, modeled as a sudden jump in $P_a$ to a new opinion value $P\neq x_i(0)$, i.e. $P_a(t)=P\,\Theta(t-t_0)$. Here, $\Theta(t)$ denotes the Heaviside step function. Without loss of generality, we initialize the system at $x_i(0)=0$ and assume that the bias $P_a$ increases at $t_0$ to a positive value $P>0$. Such dynamics are exemplarily depicted in Fig.~\ref{fig:fig1}(a). 

To quantify the coherence of opinions during the transient towards a new consensus, we define the following coherence measure
\begin{equation}\label{eq:disturbance-measure-opinion-leadership}
 \mathcal{C}(a)=\sum_i \int\limits_{0}^\infty|{x_a}(t)-{x_i}(t)|{\mathrm d}t\,.
\end{equation}
Equation~\eqref{eq:disturbance-measure-opinion-leadership} integrates the opinion distances of all agents to the opinion of the stubborn agent $x_a$ over time. During the crossover starting at $t=0$ the ensemble is driven from  $x_i(0)=0$ to a new consensus value given by the final bias magnitude of agent $a$, i.e. $x_i(t\rightarrow\infty)=P\,\, \forall\, i$.  By definition, the opinion coherence decreases with increasing value of $\mathcal{C}$. 

An expression for $\mathcal{C}$, depending on the index $i=a$ of the stubborn agent, can be derived by inserting Eq.~\eqref{eq:expansion_coefficients} into Eq.~\eqref{eq:spec-decomp}. It yields the general equation for the opinion of agent $i$,
\begin{eqnarray}\label{eq:traj}
x_i(t)=\kappa P \sum_\alpha \frac{u_{\alpha,a}^{(\kappa)}}{\lambda_\alpha^{(\kappa)}}(1-e^{-\lambda_\alpha^{(\kappa)} (t-t_0)})\,u_{\alpha,i}^{(\kappa)} \;\;,
\end{eqnarray}
for $t>t_0$, in the case of a single stubborn agent.
Taking the limit $t\rightarrow\infty$ and using Eq.~(\ref{eq:1/k}) yields,
\begin{eqnarray}\label{eq:traj_infty}
x_i(t\rightarrow \infty)=\kappa P \sum_\alpha \frac{u_{\alpha,a}^{(\kappa)} u_{\alpha,i}^{(\kappa)}}{\lambda_\alpha^{(\kappa)}}=P \; ,
\end{eqnarray}
which ensures the emergence of a new consensus at $x_i(t\rightarrow\infty)=P$ as final state of the system. Integrating Eq.~(\ref{eq:traj}), $\mathcal{C}$ can be expressed as
\begin{align}\label{eq:coherence}
\mathcal{C}(a)&=-\kappa P\sum_i\sum_\alpha\frac{{u_{\alpha,a}^{(\kappa)}}^2-u_{\alpha, a}^{(\kappa)} u_{\alpha,i}^{(\kappa)} }{{\lambda_\alpha^{(\kappa)}}^2} \; ,
\end{align}
yielding a closed form expression for the opinion coherence during a consensus change. In subsection~\ref{subsec:numerical-consensus-change} Eq.~\eqref{eq:coherence} will be utilized to rank the nodes of different networks with respect to the value of $\mathcal{C}$. We remark that, for more than one stubborn agent, i.e. $n_s>1$, with the same bias, the coherence measure of the system is given by Eq.~(\ref{eq:coherence}), where one has to additionally sum over all biased agents $a\in V_s$\,. Note that the system size $(n, n_e)$ is encoded in the spectrum of $\mathbb{L}^{(\kappa)}$ and therefore implicitly enters the definition of $\mathcal{C}$ in Eq.~\eqref{eq:coherence}. An explicit normalization of $\mathcal{C}$ might nevertheless be found useful to compare differently sized systems. However, such an additional normalization is omitted in the present study, since we focus on systems with equal or comparable sizes.

\section{Opinion heterogeneity}\label{sec:two-opposing-leaders}
In contrast to the previous section we now discuss cases in which the final stable state of the system is not characterized by a full consensus. Within the Taylor model, such non-consensus states arise only for multiple differently biased stubborn agents.
In the following, we consider the simplest setup of two antagonistic stubborn agents, i.e. $V_s=\{a,b\}$. Furthermore, we assume that their biases are perfectly balanced, i.e. $P_a(t)=P\Theta(t-t_0)=-P_b(t)$, to neglect effects stemming from different bias magnitudes ($|P_a|\neq |P_b|$).
This highly idealized situation allows to relate the influence of the network structure and the placement of biased agents therein to the characteristics of the emerging opinion states. 

As depicted in Fig.~\ref{fig:fig1}(b), the described setup indeed results in a heterogeneous opinion distribution, deviating strongly from a full consensus. Starting from an initial consensus, i.e. $x_i(0)=P_a(0)=P_b(0)=0\,\,\forall\,i$, the stubborn agents develop at $t=t_0$ opposite biases of magnitude $P$, towards positive ($a$) and negative ($b$) opinions.

The general time evolution of opinions in the system is formally solved by
\begin{align}\label{eq:general-solution-two-leaders}
\begin{split}
{x_i}(t)=\kappa P\sum_\alpha\frac{u_{\alpha,a}^{(\kappa)} - u_{\alpha,b}^{(\kappa)}}
{\lambda_\alpha^{(\kappa)}}(1-e^{-\lambda_\alpha^{(\kappa)} (t-t_0)}) {u}_{\alpha,i}^{(\kappa)}\;\;, t>t_0\;.
\end{split}
\end{align}
This yields the starting point for the considerations below. In the following we analytically quantify different properties of emerging non-consensus states in the case of two antagonistic stubborn agents.

\subsection{Opinion association}\label{sec:leader-association}
In the case of two antagonistic stubborn agents, we are interested in which of the two will have the greater impact on the population. More specifically in terms of opinion dynamics, we ask: 
which stubborn agent influences the majority -- or certain relevant agents in favor of his own opinion?
Similar questions arise for influence maximization in social systems, which usually revolve around viral marketing and information dissemination \cite{chen2009efficient}.
Here, we investigate the problem from a different angle and take into account competing effects of two opposed sources of influence. Similar questions were previously studied for systems of discrete opinion states \cite{yildiz2011discrete,masuda2015opinion}. 
We expect that an arbitrary agent is associated to the stubborn agent to whom he has the smaller opinion distance. Based on these assignments we investigate which of the two biased agents is able to influence most agents in the system.
Real-world correspondences of the framework might include presidential elections or political referendums. Rather than their precise opinion, the rough orientation of agents towards one or the other opinion stance might determine e.g. their voting behavior. Our graph-theoretic approach, presented below, yields an efficient tool to investigate this type of binary social influence on a network level. 

Without loss of generality we initialize the system in a state of full consensus at $x_i(0)=0$. At time $t_0$ the two stubborn agents develop antagonistic biases of magnitude $P$ as described previously. Hence, the opinion of each agent $i$ is generally shifted either to the positive or negative side of the opinion scale, i.e. $x_i(t\rightarrow\infty) \lessgtr 0$.  Due to the symmetric arrangement of the final opinions of stubborn agents around $x=0$ [see Eq.~(\ref{eq:final-state-x_i-two-leaders}) below] this yields a shorter opinion distance to the positively (negatively) biased stubborn agent $a$ ($b$). 
In the following we illustrate this association problem within the introduced Laplacian formalism. 

According to Eq.~\eqref{eq:general-solution-two-leaders}, the limit $t\rightarrow\infty$ yields the final opinion of an arbitrary agent $i$ as
\begin{align}
\begin{split}\label{eq:final-state-x_i-two-leaders}
x^{\infty}_i&=\kappa P \sum_{\alpha}\frac{u_{\alpha,a}^{(\kappa)} - u_{\alpha,b}^{(\kappa)}}{\lambda_\alpha^{(\kappa)}}
u_{\alpha,i}^{(\kappa)}.
\end{split}
\end{align}
Together with the definition of MRDs in Eq.~(\ref{eq:eig}) we get
\begin{align}
\begin{split}\label{eq:final-state-x_i-two-leaders}
x^{\infty}_i=\frac{\kappa P}{2}[\Omega^{(\kappa,1)}_{bi}(\{a,b\})-\Omega^{(\kappa,1)}_{ai}(\{a,b\})] \;
\end{split}
\end{align}
(see \ref{app:leada} for derivation). Remarkably, Eq.~\eqref{eq:final-state-x_i-two-leaders} suggests that it is possible to reformulate the problem of opinion association, originally defined in opinion space, in terms of MRDs. Instead of computing opinion distances in the final state, one can equivalently evaluate the involved MRDs based on $\mathbb{L}^{(\kappa)}$. Hence, to determine the association of an agent $i$ to one of the two biased agents, it suffices to compute the sign of Eq.~\eqref{eq:final-state-x_i-two-leaders}. Note that in highly symmetric networks a subset of agents may have identical opinion distances to both agents $a$ and $b$ and therefore stay undecided. Such cases are captured by the formalism resulting in $\Omega^{(\kappa,1)}_{ai}=\Omega^{(\kappa,1)}_{bi}$.

Albeit the presented formalism is rather simplistic, it yields insights into the emerging opinion formation in cases of two competing opinion camps, represented by a pair of antagonistically biased agents. In Sec.~\ref{sec:num} we will apply it to different exemplary networks and interpret the results in terms of MRDs.

\subsection{Opinion heterogeneity}\label{sec:pol}
Going beyond the high level description of opinion associations, we now aim to take a closer look at the emerging non-consensus states. While the association problem was introduced in the context of binary opinion assignments, specific properties of the final heterogeneous opinion states were ignored. In the following we characterize those and relate them to the influence network and the position of stubborn agents.

We focus our discussion on three descriptors, which quantify $(i)$ the distance between the most extreme opinions in the system, $(ii)$ the opinion variance of the ensemble as well as $(iii)$ the final mean opinion. While the former two measures capture heterogeneity of opinions, we quantify, using the latter, how strongly the opinion of one stubborn agent is favored over the other by the population of agents.

In the case of a pair of differently biased stubborn agents, the two most extreme opinions in the system are generally taken by those. Hence, to determine the maximum spread of opinions, it suffices to consider the final opinion distance between the two stubborn agents. It is defined as $D_\mathrm{max}=x_a^\infty-x_b^\infty$. 
Using Eq.~(\ref{eq:final-state-x_i-two-leaders}) $D_\mathrm{max}$
can be reformulated as (see \ref{app:DL} for the derivation)
\begin{align}
\begin{split}
D_\mathrm{max}(\{a,b\})
=\kappa P\sum_\alpha\frac{(u_{\alpha,a}^{(\kappa)} 
- u_{\alpha,b}^{(\kappa)})^2}{{\lambda_\alpha^{(\kappa)}}} =\kappa P \Omega_{ab}^{(\kappa, 1)}. \label{eq:theory-leader-distance}
\end{split}
\end{align}
As in the case of the opinion association problem, the relevant opinion distance ($D_\mathrm{max}$) can be expressed in terms of the corresponding MRD. Note that in contrast to the case discussed in Sec.~\ref{subsec:opinion-coherence}, here, the opinions of agents do generally not reach their final biases $P_a$ and $P_b$, respectively.
Instead,  $D_\mathrm{max}$ is generally smaller than the difference of bias opinions, i.e. $D_\mathrm{max}\leq2P$. In \ref{app:stub} we show that the equality holds in the limit of an infinite stubbornness, i.e. $\kappa\rightarrow\infty$\,.

In \ref{app:mu_sigma} we demonstrate the derivation for the mean $\mu_x=n^{-1}\sum_i x_i$ and the variance $\sigma_x^2=n^{-1}\sum_i(x_i-\mu_x)^2$ of the final opinion distribution, which can also be expressed in terms of MRDs. This results in the following expressions 

\begin{equation}\label{eq:opinion_mean}
    \mu_x(\{a,b\}) =\frac{\kappa P}{2}\left[ C_1^{-1}(b) - C_1^{-1}(a) \right]
\end{equation}
and 
\begin{equation}\label{eq:opinion_variance}
\sigma^2_x(\{a,b\}) = \left(\frac{\kappa P}{2}\right)^2 \Bigg(\frac{4\Omega^{(\kappa,2)}_{ab}}{n}-\left[ C_1^{-1}(b) - C_1^{-1}(a) \right]^2\Bigg)\,.    
\end{equation}
The second order MRD, $\Omega_{ab}^{(\kappa, 2)}$, and the centrality $C_1$, were both defined in Sec.~\ref{sec:resistance_distance}. Note, that the first term in Eq.~(\ref{eq:opinion_variance}) can be also expressed by first order MRDs as $\Omega_{ab}^{(\kappa, 2)}=\frac{1}{4}\sum_i\left( \Omega_{bi}^{(\kappa,1)}-\Omega_{ai}^{(\kappa,1)} 
\right)^2$. Hence, all three introduced quantities presented in Eqs.~\eqref{eq:theory-leader-distance}-\eqref{eq:opinion_variance} can be formulated solely in terms of $\Omega_{ij}^{(\kappa,1)}$. These analytical results directly relate properties of the final heterogeneous state of opinions to properties of the  influence network structure and will be discussed more thoroughly in the following.

\section{Numerical Results}\label{sec:num} In this section we investigate the influence of stubborn agents on different network topologies. In the case of a single stubborn agent, we focus on the transient dynamics towards a new consensus, cf. Sec.~\ref{subsec:opinion-coherence}. For two antagonistically biased stubborn agents (Sec.~\ref{sec:two-opposing-leaders}) we first illustrate the opinion association framework. In a second step we discuss the emerging opinion heterogeneity. 

To establish a connection between the topology of the influence network and the impact of stubborn agents, we systematically evaluate the derived quantities, defined in Eq.~\eqref{eq:coherence} and Eqs.~\eqref{eq:theory-leader-distance}-\eqref{eq:opinion_variance} on Watts-Strogatz (WS) and stochastic block (SBM) model networks. While the WS model allows to draw conclusions about the influence of small-worldness (i.e. short average path length and high clustering coefficient), SBM networks are used to probe the effects of community structure on the opinion formation process.
Watts-Strogatz networks are discussed in terms of the link rewiring probability $p_r$.
Initially, each node is connected to its $K_\mathrm{WS}$ nearest neighbors. Increasing values of $p_r$ change this highly clustered configuration of a regular ring lattice gradually into a random network, while the average shortest path length is drastically reduced \cite{watts1998collective}.
In the case of the SBM we assign the nodes of the network to two equally sized groups of size $n/2$. Based on this partition we vary the probability for links within a group, $p_\mathrm{intra}$, versus the probability for connections between nodes in two different groups ($p_\mathrm{inter}$). For high values of $p_\mathrm{intra}$ (and low $p_\mathrm{inter}$) this leads to a pronounced community structure.
To exclude effects stemming from different system sizes, we fix the number of nodes and edges to $n=50$ and $n_e=200$, where not differently indicated. In order to fix the expected number of  edges of the discussed SBM networks to $\langle n_e\rangle=200$ we implement a linear relation between $p_\mathrm{inter}$ and $p_\mathrm{intra}$ (cf. \ref{app:network-models-dataset}). 
While increasing $p_\mathrm{intra}$ on the interval $p_\mathrm{intra}\in[0.2,1/3[$, the inter-group link probability is decreased accordingly.
Note that, by construction, the network disintegrates in two disconnected communities for $p_\mathrm{intra}\rightarrow 1/3$, cf. \ref{app:network-models-dataset}.
Therefore, we merely approach this value in our analysis to ensure connected influence networks.

In addition to the two synthetic network models (WS, SBM) we briefly discuss the case of a friendship graph of highschool students \cite{nr_data} as an application of our theoretical framework to an empirical network. Here, we focus on the case of two antagonistically biased agents and investigate how the heterogeneity of opinion states changes upon randomizing the original network structure. For details on the empirical data set and the synthetic network models, see \ref{app:network-models-dataset}. 


Unless otherwise stated, we set the stubbornness $\kappa$ and the bias magnitudes $P$ of biased agents to unity, i.e. we have $\kappa = 1$ and $P_a(t)=\Theta(t-t_0)$ in the case of a single stubborn agent and  $P_a(t)=\Theta(t-t_0)=-P_b(t)$ for two opposed stubborn agents. 


\subsection{Consensus change}\label{subsec:numerical-consensus-change}
In the following, we consider the setup as introduced in Sec.~\ref{subsec:opinion-coherence}. Starting from the initial state $x_i=0$ $\forall i$\,, a single stubborn agent drives the system to a new consensus value given by his/her final bias $P=1$. 

\begin{figure}[h!]
\includegraphics[width=0.99\linewidth]{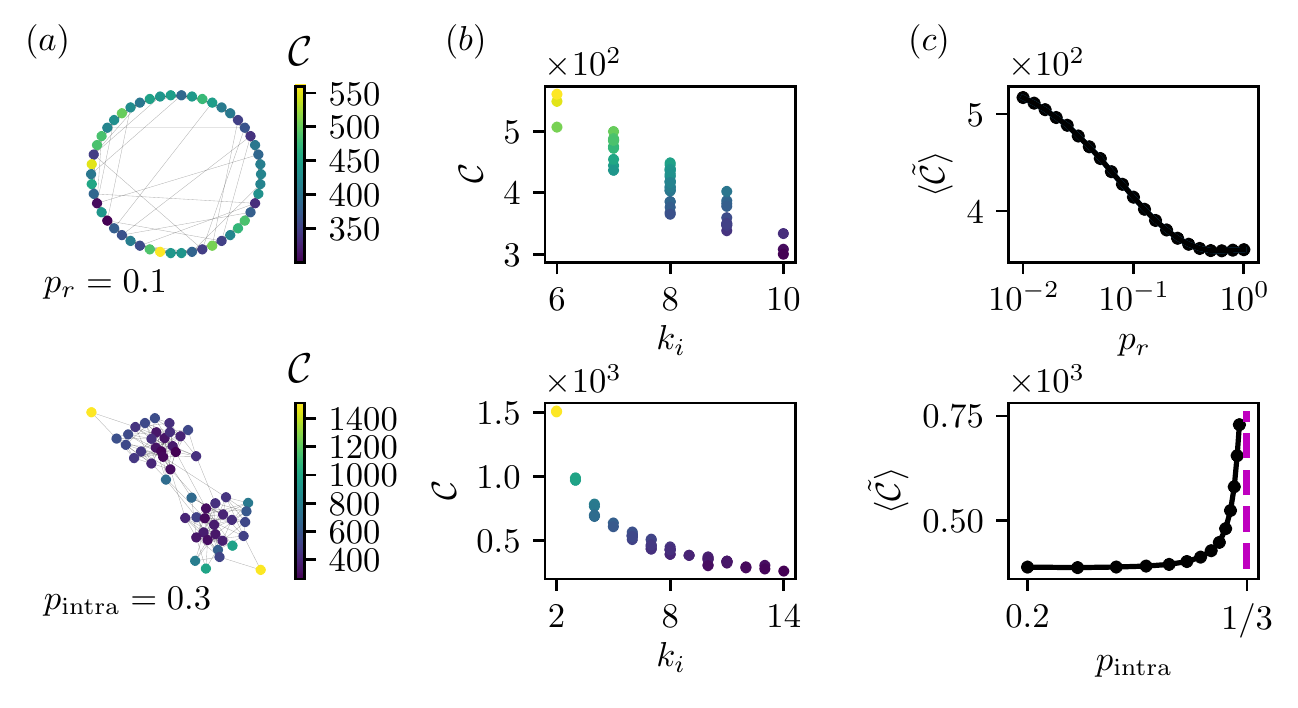}
\caption{Opinion coherence in the case of a single stubborn agent. Panel (a): the color of a node $i$ depicts the resulting value of $\mathcal{C}(i)$ for a WS network (top) and a SBM network (bottom). In panel (b) the values of $\mathcal{C}(i)$ are plotted against the degrees of the corresponding nodes in the WS (top) and SBM (bottom) networks. Panel (c) shows $\tC$ as a function of $p_r$ (top) and $p_\mathrm{intra}$ (bottom) for WS and SBM networks, respectively. The dashed vertical line is placed at $p_\mathrm{intra}=1/3$, where the SBM network disintegrates in two disconnected communities. The system parameters were set to: $n=50$, $n_e(\langle n_e\rangle)=200$ for WS (SBM) networks. Note that $\langle \tilde{\mathcal{C}}\rangle$ denotes the average of $\tC$ over $5000$ networks.}
\label{fig:fig_coherence_map}
\end{figure}

First, we evaluate the resulting coherence measure $\mathcal{C}(i)$ depending on the node  $i$ on which the biased agent resides. This is depicted in Fig.~\ref{fig:fig_coherence_map}(a) (color code) for a WS graph (top) and a SBM network (bottom), respectively. Increasing values of $\mathcal{C}$ are depicted in brightening colors from dark blue to yellow. A closer look at the network illustrations suggests that the brightness of a node decreases with its degree. Especially in the case of the SBM network, which is depicted using a  force-directed algorithm \cite{fruchterman1991graph}, we observe that some peripheral nodes with lower degree are depicted in brighter colors compared to more central nodes of the same community.

This is confirmed in Fig.~\ref{fig:fig_coherence_map}(b), where the opinion coherence $\mathcal{C}(i)$ is plotted against the degree of node $i$. For both networks, WS (top) and SBM (bottom), we observe that the quantities are negatively correlated. As the degrees increase the values of $\mathcal{C}$ decrease. This suggests, that in order to maximize network opinion coherence during a consensus change, it is advantageous to place stubborn agents on nodes of rather high degree. Note, however, that $k_i$ cannot yet be utilized as an unambiguous predictor for $\mathcal{C}$. Clearly, for nodes of different degree, the corresponding values for $\mathcal{C}$ overlap, especially in the case of the WS network. Interestingly, this effect becomes less pronounced for increased edge densities, as shown in Fig.~\ref{fig:fig8_App_physA}(a) of \ref{app:App_merge}, where we depict the relationship of Fig.~\ref{fig:fig_coherence_map}(b) for a WS network of $n=50$ and $n_e=500$. 


The observed variability of $\mathcal{C}$ suggests, that the consideration of a specific stubborn agent is not sufficient to systematically investigate opinion coherence on a network level. To circumvent this limitation, all nodes need to be incorporated in our formalism. Hence, we introduce a network measure for opinion coherence
\begin{eqnarray}\label{eq:cohs}
\tilde{\mathcal{C}}=\langle \mathcal{C} \rangle_{\{ a \}} \; ,
\end{eqnarray}
defined as the average of Eq.~\eqref{eq:coherence} over all nodes in the network. 
Fixing $n$ and $n_e$ for WS and SBM networks, we discuss in the following, the change of $\tilde{\mathcal{C}}$ in terms of $p_r$ (WS) and $p_\mathrm{intra}$ (SBM). The results are depicted in Fig.~\ref{fig:fig_coherence_map}(c). 

Note that, as the SBM and WS models generate random graphs, we consider $\tC$ in terms of an additional average, over many such network realizations. For the ease of notation we omit this second average in the text but indicate it in Fig.~\ref{fig:fig_coherence_map} as $\langle \tilde{\mathcal{C}} \rangle$. The same applies to the network-averaged quantities in the case of two stubborn agents, discussed in the next section.

In the case of WS networks the coherence measure decreases rapidly from its maximum value at $p_r=0$ as the network is randomized. Note the logarithmic scale of the $x$--axis. For larger values of $p_r$ it decreases more slowly and approaches a constant value, cf. [Fig.~\ref{fig:fig_coherence_map}(c, top)]. This trend of $\tilde{\mathcal{C}}$ as a function of the rewiring probability is in line with previous findings on consensus dynamics in small-world networks. In Ref.~\cite{li2006dynamics} it was found, using a model of local majority updating, that the time to consensus monotonically decreases with the rewiring probability $p_r$. In our case, the monotonic decrease of $\tilde{\mathcal{C}}$ as a function of $p_r$ can be related to the decrease of MRDs between the single stubborn agent and the remaining agents. As we will additionally demonstrate below for settings of two opposed stubborn agents, smaller MRDs increase the overall influence of an agent on the collective opinion formation process.
In Fig.~\ref{fig:fig_coherence_map} (c, bottom) we depict the values of $\tC$ on different configurations of SBM networks. For low and moderate values of $p_\mathrm{intra}$ we find that the opinion coherence remains rather constant. As $p_\mathrm{intra}$ is further increased, $\tilde{\mathcal{C}}$ diverges. By construction, this divergence is precisely located at $p_r=1/3$ for which the network disintegrates into two disconnected sub graphs. 
Note that the SBM network depicted in the bottom panel of Fig.~\ref{fig:fig_coherence_map}(a), was generated for $p_\mathrm{intra}=0.3<1/3$. Hence, the probability for links between both groups $p_\mathrm{inter}$, is still finite and therefore the two communities in the resulting network are connected.
For $p_\mathrm{intra}\geq1/3$ instead, the respective opposed community is isolated from the influence of a single stubborn agent, which results in the divergence of $\mathcal{C}$\,. Indeed, half of the population will remain at the initial consensus at $x_i=0$, cf. Eq.~\eqref{eq:disturbance-measure-opinion-leadership}.
Within our formalism, this limiting case can be understood in terms of modified resistance distances. The MRDs between the single stubborn agent and the agents in the opposed (disconnected) subgraph diverge. As Eq.~\eqref{eq:final-state-x_i-two-leaders} for two stubborn agents, $V_s=\{a,b\}$, suggests, the influence of a stubborn agent $i=a$ onto any other agent in the system $j$ decreases as $\Omega^{(\kappa, 1)}_{aj}$ increases. This effect leads to an increasing opinion diversity during the consensus change for $p_\mathrm{intra}\rightarrow 1/3$. This finding is in line with previous results on consensus dynamics on networks exhibiting a pronounced community structure. In \cite{shin2010tipping} it was shown, using a simple model for information accumulation, that a decrease of inter-community connectivity hampers the establishment of a global consensus.

\subsection{Opinion heterogeneity}
\subsubsection{Opinion association}
\begin{figure}
\includegraphics[width=0.99\linewidth]{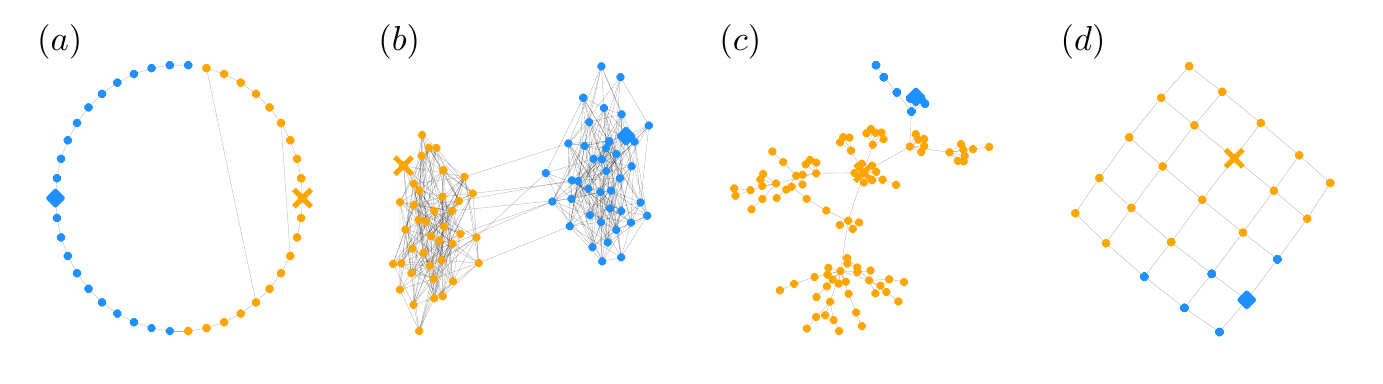}
\caption{Opinion association to pairs of antagonistically biased agents.
The nodes of the two stubborn agents are depicted as orange crosses ($i=a$) and blue squares ($i=b$), respectively. Given their positions, i.e. the set of stubborn agents $V_s$, our formalism allows to efficiently determine the association of each agent as the sign of Eq.~\eqref{eq:final-state-x_i-two-leaders}. Each agent is colored according to its opinion association. The subpanels depict results for different example networks: WS (a), SBM (b), Barab\'asi-Albert (BA) \cite{barabasi1999emergence} (c) and a lattice network 
(d).}\label{fig:leader-follower-association}
\end{figure}

As the derivation in Sec.~\ref{sec:leader-association} reveals, the opinion association problem can be analytically treated in terms of the modified Laplacian $\mathbb{L}^{(\kappa)}$. To demonstrate the applicability of the framework, we evaluate Eq.~\eqref{eq:final-state-x_i-two-leaders} on four different synthetically generated networks. As we will see in the following, our formalism allows to understand the associations of agents in terms of MRDs.
In Fig.~\ref{fig:leader-follower-association}, the positively (negatively) biased stubborn agent is depicted as orange cross (blue diamond). The remaining agents are depicted as dots, where the color displays their final association to one of the two stubborn agents.

In Fig.~\ref{fig:leader-follower-association}(a) two stubborn agents are placed on opposite sides of a WS network. Even though the MRDs between agent $i=a$ (orange cross) and the remaining nodes on the right half part of the graph have been reduced by the random edge rewiring, the MRDs to the left half part of the cycle have merely been changed. This is due to the fact that no shortcuts were introduced to directly connect one of the stubborn agents to the opposite side of the graph. Consequently, the opinion association is clearly split into two separated parts of roughly equal size. 
Next we illustrate the case of an influence network generated by the SBM with two densely connected regions of equal size. Here, we assume that one stubborn agent is placed within each of the communities. This setup leads most probably to a situation depicted in Fig.~\ref{fig:leader-follower-association}(b). Intuitively, the high internal connectivity within each community leads to a drastic reduction of MRDs of intra- versus inter-community node pairings. Therefore, a stubborn agent placed within a specific community has a high chance of attracting all agents within that group of nodes. 
While the previous two networks (WS, SBM) obeyed rather narrow degree distributions we now turn to the case of a scale-free Barab\'asi-Albert (BA) network \cite{barabasi1999emergence} shown in Fig.~\ref{fig:leader-follower-association}(c). We find that the majority of agents is associated to the positively biased agent (orange cross), which is placed on the node with the highest degree. By contrast, the negatively biased agent, placed on a peripheral node (blue square), merely attracts agents in its close vicinity. As Eq.~\eqref{eq:final-state-x_i-two-leaders} suggests, the opinion association of an arbitrary agent is determined by the minimal MRD to one of the stubborn agents. Hence, the average MRD of a stubborn agent to all other nodes $j$, $\langle \Omega^{(\kappa,1)}_{ij}\rangle_j$, crucially determines his/her influence on the population. Due to the increased connectivity of high degree nodes, this mean MRD is generally decreases for such nodes and leads to higher centrality values as defined in Eq.~\eqref{eq:centr}. Hence, stubborn agents placed on high degree nodes, so-called hubs, are expected to attract more agents than low degree agents at the periphery of the network. 
Finally, we depict a two-dimensional lattice network in Fig.~\ref{fig:leader-follower-association}(d). Compared to the negatively biased agent (blue square), which is placed on one of the boundaries of the lattice, the stubborn agent with positive bias (orange cross) possesses a more central position. Its central placement, away from the boundaries, favors small MRDs to most other agents in the system. Hence, s/he is able to associate the larger number of agents.

\subsubsection{Opinion heterogeneity}
\begin{figure}[h!]
\includegraphics[width=0.99\linewidth]{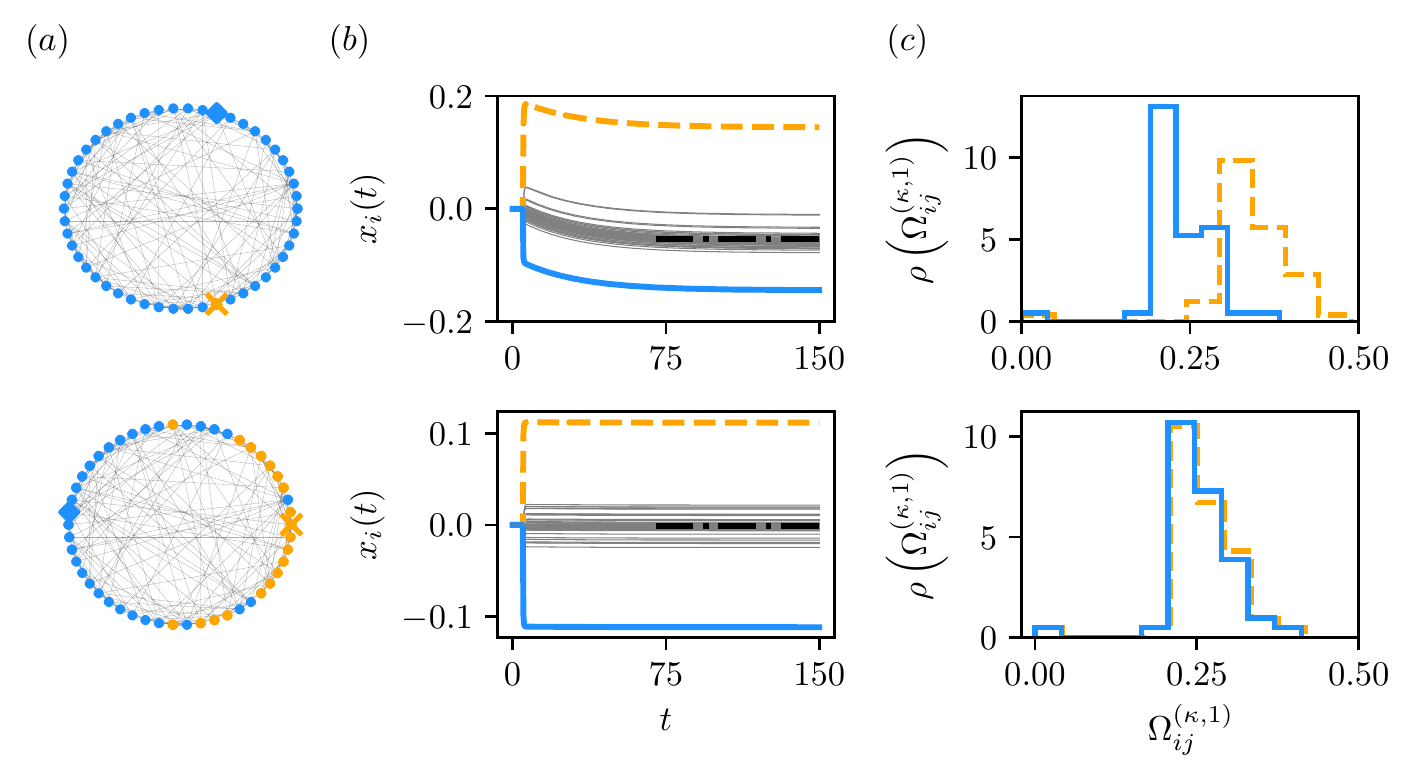}
\caption{Opinion associations, transient dynamics and final distributions of MRDs for two different sets of stubborn agents (bottom/top panels) on a fixed WS network with $p_r=0.5$. 
In panel (a) the stubborn agents are depicted as orange cross ($i=a$) and blue square ($i=b$). Small dots show the remaining agents color-coded according to their opinion association. 
Panel (b) shows opinion trajectories of the two stubborn agents depicted as thick orange dashed ($i=a$) and blue solid ($i=b$) lines. The opinions of the remaining agents are depicted in thin gray lines. The final mean opinion is shown as the black dashed dotted line. Note the different scales on the $y$--axes.
Panel (c): Distributions of MRDs for both stubborn agents $(a,b)$ to all remaining nodes $j$ in the system are shown as thick orange dashed ($i=a$) and blue solid ($i=b$) lines, respectively.}
\label{fig:two_pairs}
\end{figure}

In the case of a pair of antagonistically biased agents the system will generally not persist in or reach a state of full consensus. As predicted by Eqs.~\eqref{eq:theory-leader-distance}-\eqref{eq:opinion_variance}, important characteristics of the emerging non-consensus states crucially depend on the network's topology and the positions of the stubborn agents therein. This is demonstrated in Fig.~\ref{fig:two_pairs}, where we show opinion associations [panel (a)] resulting from two different sets of stubborn agents on a fixed network topology. As previously the orange cross (blue square) illustrates the position of the positively (negatively) biased stubborn agent. The remaining nodes are depicted as dots, colored according to their association to one of the biased agents. We first note that, despite identical network topologies, the association of opinions substantially differs between the different sets $V_s$. For the first pair of stubborn agents, depicted in the top panel of Fig.~\ref{fig:two_pairs}(a), all agents are associated to the negatively biased agent. By contrast, in the bottom panel of Fig.~\ref{fig:two_pairs}(a) we observe a more balanced situation, in which only slightly more agents are associated to stubborn agent $b$ (blue square). The final association of opinions is reflected in the transient dynamics of the system, depicted in Fig.~\ref{fig:two_pairs}(b). The opinions of stubborn agents are shown as thick dashed ($i=a$) and solid ($i=b$) lines. Thin gray lines correspond to the opinion trajectories of the remaining agents and the final mean opinion $\mu_x$ [cf. Eq.~\eqref{eq:opinion_mean}] is depicted as the thick dashed dotted line. The value of $\mu_x$ indicates the overall trend in the final opinion distribution. While $\mu_x$ is close to zero in Fig.~\ref{fig:two_pairs}(b, bottom), it is clearly shifted to a negative value for the other set of stubborn agents , cf. Fig.~\ref{fig:two_pairs}(b, top). For different sets of stubborn agents also the values of $D_\mathrm{max}$ and $\sigma^2_x$ are subject to change and decrease from the top to the bottom panel of Fig.~\ref{fig:two_pairs}(b).

Interestingly, these general trends can be related to the distributions of MRDs between the stubborn agents and the remaining agents in the population, denoted as $\rho\left(\Omega^{(\kappa,1)}_{ij}\right)$ with $i\in V_s$, cf. Fig.~\ref{fig:two_pairs}(c). 
While the distributions mostly overlap for the second set $V_s$ [Fig.~\ref{fig:two_pairs}(c) bottom], they strongly differ for the depicted case in the top panel of Fig.~\ref{fig:two_pairs}(c). 
Here, $\rho\left(\Omega^{(\kappa,1)}_{bj}\right)$ (blue solid line) is clearly shifted towards lower values with respect to $\rho\left(\Omega^{(\kappa,1)}_{ia}\right)$ (orange dashed line). This explains the collective bias of the population towards agent $b$ and is quantified by the corresponding mean MRDs for both stubborn agents $\langle\Omega^{(\kappa,1)}_{aj}\rangle_j$ and $\langle\Omega^{(\kappa,1)}_{bj}\rangle_j$,  and sets of $V_s$: $0.343$ ($i=a$), $0.236$ ($i=b$) [Fig.~\ref{fig:two_pairs}(c) top] and $0.259$ ($i=a$), $0.257$ ($i=b$) [Fig.~\ref{fig:two_pairs}(c) bottom]. 

\begin{figure*}[t]
\includegraphics[width=1.\linewidth]{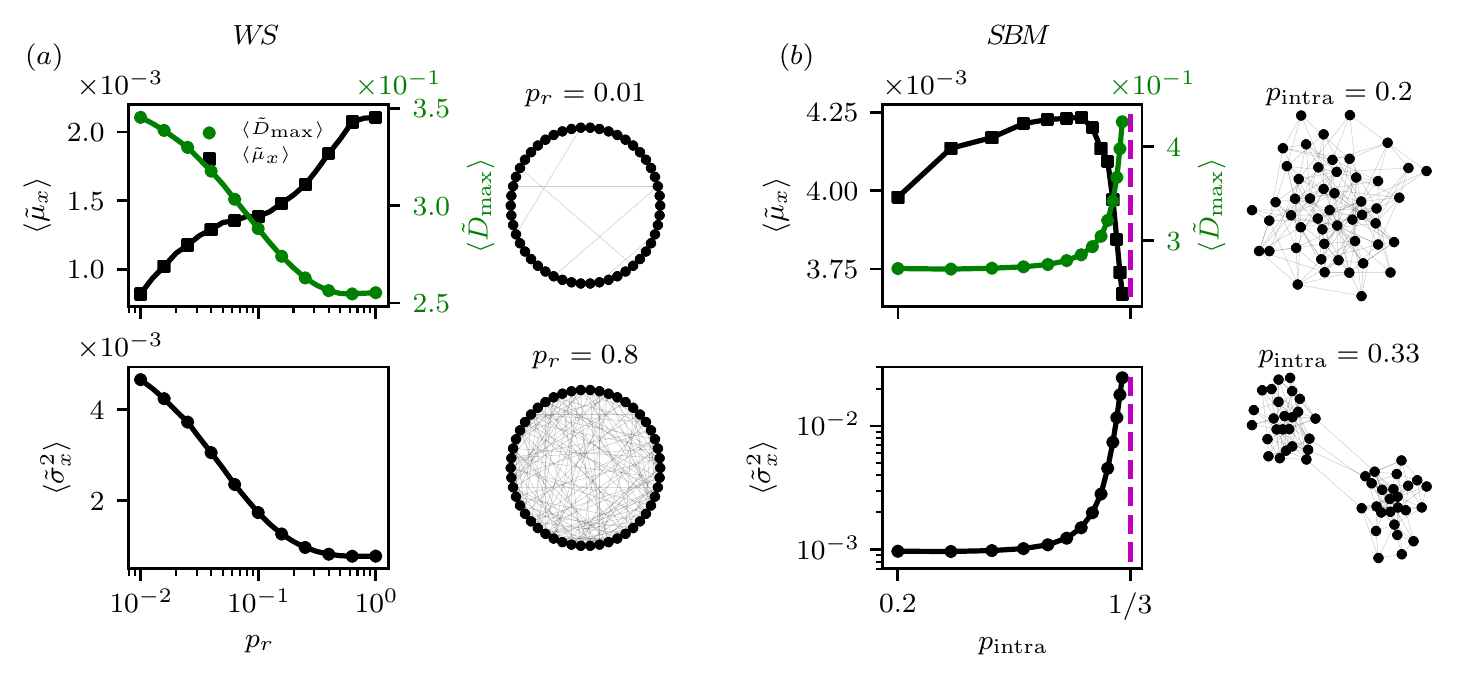} 
\caption{Maximum opinion spread, absolute mean opinion and opinion variance for WS (a) and SBM networks (b). The quantities are shown as functions of the rewiring probability $p_r$ (WS) or the probability of intra-group links $p_\mathrm{intra}$ (SBM). The number of nodes is fixed to $n=50$ and the WS networks have a constant number of $n_e=200$ edges. In the case of SBM networks the average number of edges is fixed to $\langle n_e\rangle=200$ relating the probabilities $p_\mathrm{intra}$ and $p_\mathrm{inter}$, as specified in \ref{app:network-models-dataset}. Note, that $\langle \tilde{\mu}_x\rangle$, $\langle \tilde{\sigma^2}_x\rangle$ and $\langle \tilde{D}_\mathrm{max}\rangle$ are computed as averages over $5000$ networks. The dynamical parameters of the system were chosen to be $P=P_a=-P_b=1$ and $\kappa=1$\,. To the right of panels (a) and (b) we depict two networks for each considered network model, generated for different values of the corresponding parameters $p_r$ (WS) and $p_\mathrm{intra}$ (SBM), respectively. The dashed vertical lines in panel (b) are placed at $p_\mathrm{intra}=1/3$ for which the network disintegrates into two disconnected communities.}
\label{fig:XYZ}
\end{figure*}

\begin{figure}[h!]
\includegraphics[width=0.99\linewidth]{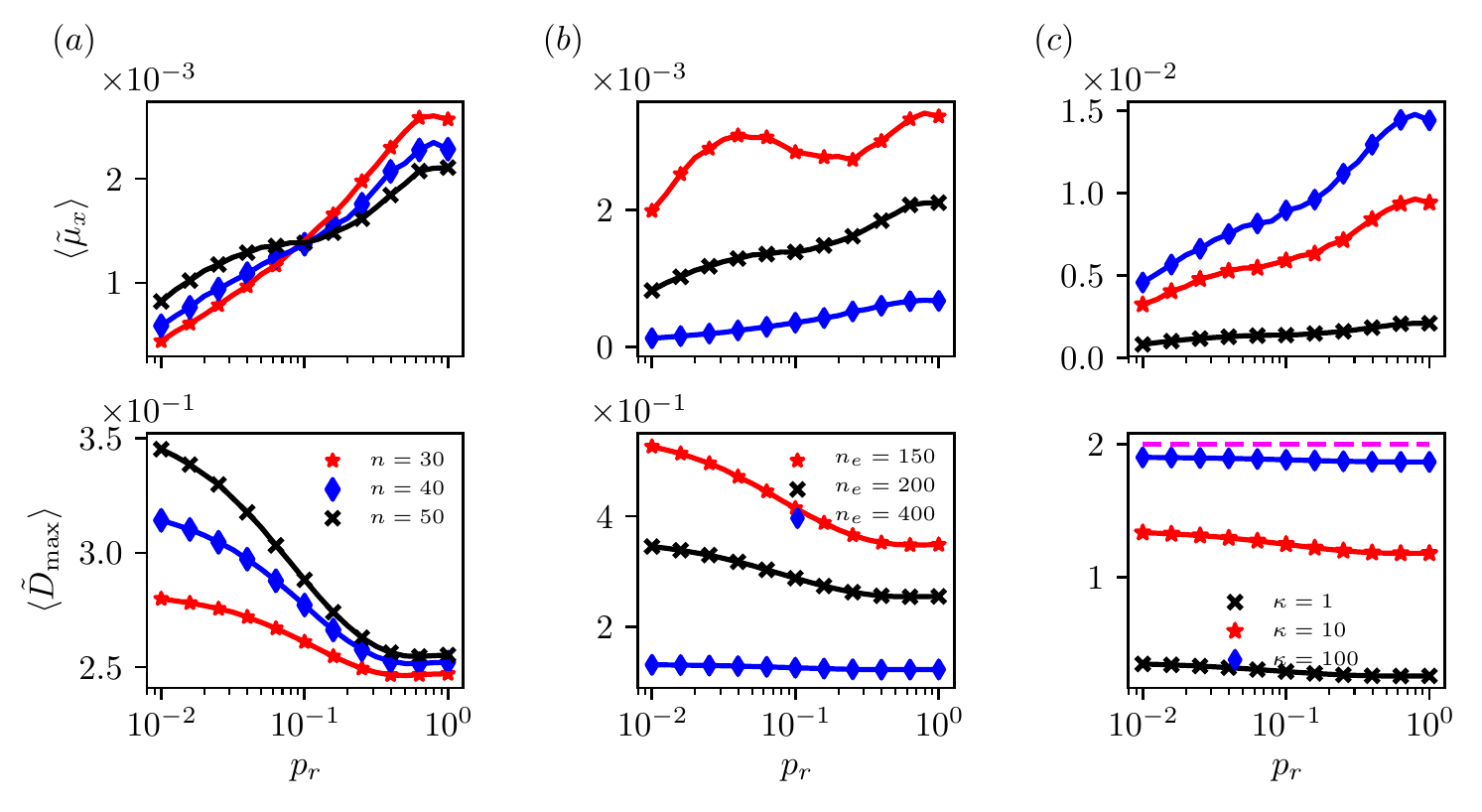}
\caption{Mean opinion $\tmu$ and maximum opinion distance $\tDL$ for WS networks and different model parameters. Panels (a) -- (c) depict the variation of a specific system parameter. Panel (a): system size with constant average degree $2n_e/n=8$. Panel (b): number of edges $n_e$ while fixing the number of nodes $n=200$. Panel (c): stubbornness parameter $\kappa$. Note, that $\langle \tilde{\mu}_x\rangle$ and $\langle \tilde{D}_\mathrm{max}\rangle$ are computed as averages over $5000$ networks.}
\label{fig:fig_param_exploration}
\end{figure}

\begin{figure}[h!]\begin{center}
\includegraphics[width=0.99\linewidth]{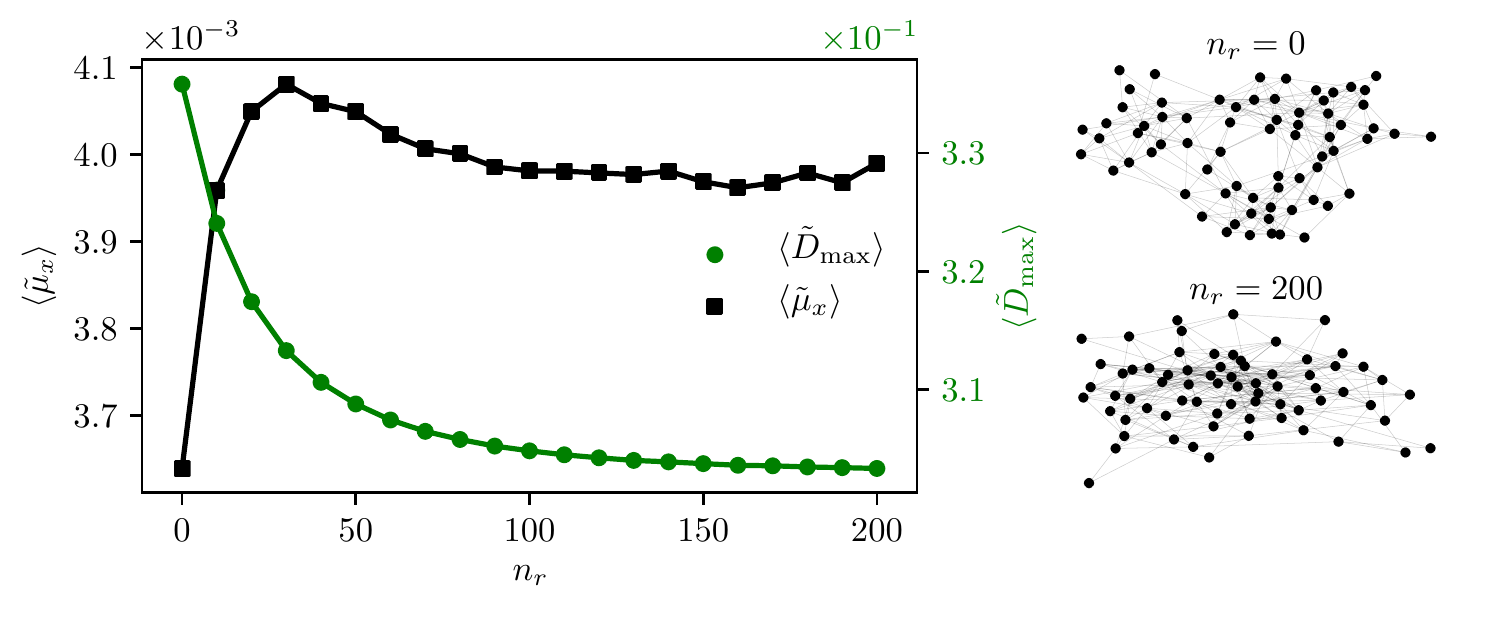}
\caption{Mean opinion value $\tmu$ and maximum opinion distance $\tilde{D}_\mathrm{max}$ for an empirical friendship network of high-school students plotted against the number of rewired edges $n_r$ of the original network (left panel). The right panels depict the original friendship network ($n_r=0$, top) and a randomized version with $n_r=200$ (bottom). Note, that for $n_r>0$, $\langle \tilde{\mu}_x\rangle$ and $\langle \tilde{D}_\mathrm{max}\rangle$ are computed as averages over $5000$ networks.}
\label{fig:fig_empirical_network_quantities}
\end{center}
\end{figure}

These findings suggest that for a pair of stubborn agents, and just as in the case of a single biased agent, the system's response critically depends on $V_s$. Therefore, to characterize the heterogeneity of opinion states on a network level, we define 

\begin{subequations}\label{eq:global-polarization-measures}
\begin{align}
\tilde{D}_\mathrm{max} &=\langle D_\mathrm{max}\rangle_{\{a,b\}}\,,\\ 
\tilde{\mu}_x &=\langle |\mu_x|\rangle_{\{a,b\}}\,,\\ 
\tilde{\sigma}^2_x &=\langle \sigma^2_x\rangle_{\{a,b\}}\,, 
\end{align}
\end{subequations}
where $\langle X\rangle_{\{a,b\}}$ denotes the average of $X$ over all possible sets $V_s = \{a,b\}$.
Those averaged quantities provide a way to investigate how the expected opinion heterogeneity depends on the parameters of a given network model. Note that, we generally consider $ \tilde{D}_\mathrm{max}$, $ \tilde{\mu}_x$, $\tilde{\sigma}^2_x$ as averages of Eqs.~(\ref{eq:global-polarization-measures}) over many realizations of randomly generated networks, i.e. $\langle \tilde{D}_\mathrm{max}\rangle$, $\langle \tilde{\mu}_x\rangle$, $\langle \tilde{\sigma}^2_x\rangle$. As in the case of $\tC$ discussed above, we omit this notation in the text but indicate it in Figs.~\ref{fig:XYZ}, \ref{fig:fig_param_exploration} and \ref{fig:fig_empirical_network_quantities}.

For large influence networks the evaluation of Eqs.~\eqref{eq:global-polarization-measures} becomes computationally expensive. This is due to the quadratic growth of possible combinations of stubborn agents with the network size in the case of $n_s=2$, demonstrated in \ref{app:App_merge}, where we compare the computational complexity for $\tDL$ (two stubborn agents) to $\tC$ (single stubborn agents).
Hence, to present exact results while restricting the time complexity of the numerical computations to a reasonable amount, we discuss networks of rather small system size of $n=50$. 
In Fig.~\ref{fig:XYZ} we evaluate Eqs.~\eqref{eq:global-polarization-measures} for different configurations of WS [panel (a)] and SBM networks [panel (b)]. Similar to Sec.~\ref{subsec:numerical-consensus-change}, we depict $\tDL$, $\tilde{\mu}_x$ and $\tilde{\sigma}^2_x$ as functions of $p_r$ (WS) and $p_\mathrm{intra}$ (SBM). Note that the smoothness of the depicted curves is due to an averaging procedure --  each symbol in Figs.~\ref{fig:fig_coherence_map}(c), \ref{fig:XYZ}, \ref{fig:fig_param_exploration} and \ref{fig:fig_empirical_network_quantities} is computed over 5000 networks with identical network parameters. 

First, we note that $\tilde{D}_\mathrm{max}$ and $\tilde{\sigma}_x^2$, behave very similarly to the coherence measure $\tC$ for both network models [Fig.~\ref{fig:fig_coherence_map}(c)]. 
For the WS model an increased rewiring probability promotes a decrease for both quantities, cf. ~Fig.~\ref{fig:XYZ}(a). According to Eq.~\eqref{eq:theory-leader-distance} the final opinion distance between the stubborn agents, $D_\mathrm{max}$, is proportional to the MRD between both agents. Due to the rewiring procedure defined by the WS model, those distances are (on average) strongly reduced, as shortcuts between different network regions are added. The effect is particularly pronounced in the small-world regime around $p_r = 0.1$. For even larger values of $p_r>0.5$ $\tilde{D}_\mathrm{max}$ slowly saturates and approaches a constant value which is expected for a random network with the same number of nodes and edges. 
Arguments along these lines also motivate the decrease of the opinion variance as a function of $p_r$. As Eq.~\eqref{eq:opinion_variance} suggests, $\tsigma$ can be expressed as the difference between a term proportional to the second order MRD between both stubborn agents and the squared opinion mean. Again, the rewiring procedure reduces the mean value of $\Omega^{(\kappa,2)}_{ab}$, while $\tmu$ increases as discussed in the following.

The dependence of $\tilde{\mu}_x$ on $p_r$ shows a qualitatively different behavior. 
Due to the perfectly symmetric structure of the ring lattice (for $p_r=0$), the centralities of both stubborn agents, $C_1(a)$ and $C_1(b)$, will generally be equal. As Eq.~\eqref{eq:opinion_mean} suggests, this results in a vanishing final mean opinion. Note however, that, although $\tilde{\mu}_x=0$, the population will not prevail in a perfect consensus. Instead,  due to the inherent symmetry of the influence network, the agents will generally split in two equally sized groups -- each half associated to one of the stubborn agents. This is suggested by the finite values of $\tilde{D}_\mathrm{max}$ and $\tilde{\sigma}^2_x$ for $p_r=0$. 
The introduction of shortcuts causes $\tilde{\mu}_x$ to rise monotonically. In contrast to $\tDL$ and $\tilde{\sigma}^2_x$, it reaches its maximum value for $p_r\rightarrow 1$.

In Fig.~\ref{fig:XYZ}(b) we depict the results for networks generated from the SBM.  
For increasing values of $p_\mathrm{intra}$ the values of both $\tDL$ and $\tilde{\sigma}_x^2$ monotonically increase. As in the former case of the WS network, their functional shape is very similar to the one of $\tilde{\mathcal{C}}$ [see Fig.~\ref{fig:fig_coherence_map}(c)]. 
While the intra-group connection probability is increased, the probability for edges between the emerging communities decreases. This leads to larger average MRDs between two randomly chosen stubborn agents, which causes $\tDL$ to increase. An analogous argument for second order MRDs between the pair of stubborn agents, i.e. $\Omega^{(\kappa,2)}_{ab}$, explains the increase of $\tilde{\sigma}^2_x$ as a function of $p_\mathrm{intra}$, cf. Eq.~\eqref{eq:opinion_variance}.

On SBM networks the absolute final mean opinion, $\tilde{\mu}_x$, shows a very  different behavior. After an initial increase to the maximum value, it rapidly decreases as $p_{\rm intra} \rightarrow 1/3$. In this limit the community structure of the network becomes very pronounced, i.e. the system approaches a limit in which only nodes within one community are connected.
This steep decrease of $\tilde{\mu}_x$ for very pronounced community structures, i.e. $p_r\rightarrow 1/3$, can be explained by cases similar to the one depicted in Fig.~\ref{fig:leader-follower-association}(c). Here, all nodes of a community are associated to the stubborn agent in the respective group of nodes. Regarding the total population, this leads to a balanced situation with respect to $\tilde{\mu}_x$ and, simultaneously, to a strong increase in $\tilde{D}_\mathrm{max}$ and $\tilde{\sigma}_x^2$ as discussed above.

In Fig.~\ref{fig:XYZ} crucial system parameters were held constant. However, in particular, the values of $n$, $n_e$ and $\kappa$ are expected to influence the opinion formation process. In the following, we therefore discuss effects arising from different system sizes and investigate the influence of increasingly stubborn biased agents. In contrast to the bias magnitude $P$, which merely appears as a prefactor in the derived quantities [cf. Eq.~\eqref{eq:coherence} and Eqs.~(\ref{eq:theory-leader-distance}-\ref{eq:opinion_variance})], the stubbornness parameter enters the definition of $\mathbb{L}^{(\kappa)}$. Therefore $\kappa$ might alter the role of biased agents in a non-trivial way. 
To investigate those aspects, we explore the behavior of $\tmu$ and $\tDL$ on WS networks for different values of $n$, $n_e$ and $\kappa$. The results are shown in Fig.~\ref{fig:fig_param_exploration}.
In panel (a) we vary the total number of nodes $n\in[30, 40, 50]$ and edges $n_e\in[120, 160, 200]$, while the average degree ($2n_e/n$) is fixed. The results depicted as black cross symbols correspond to the parameter setting investigated in Fig.~\ref{fig:XYZ}(a). For a decreasing number of nodes and edges and large rewiring probabilities the averaged mean final opinion increases. Interestingly, this effect is inverted 
for $p_r<0.1$. 
Instead, $\tDL$ is generally reduced for smaller system sizes, cf. bottom panel of Fig.~\ref{fig:fig_param_exploration}(a). This can be attributed to the overall larger average MRD between any pair of nodes for larger systems. 
In panel (b) we focus on effects arising from different link densities in the network. We therefore fix the number of nodes $n$ and vary the number of edges $n_e$. Clearly, decreasing the number of edges promotes higher values of $\tmu$. Furthermore, we find that the shape of $\tilde{\mu}_x$ changes, showing a pronounced local maximum for $n_e=150$ and low $p_r$. The edge density also affects $\tDL$. Here, in contrast to Fig.~\ref{fig:fig_param_exploration}(a), larger system sizes (now in terms of $n_e$) lead to smaller values of the maximum opinion distance. Due to the increased number of edges (for constant $n$) the MRDs between any pair of stubborn agents is decreased. This leads, according to Eq.~\eqref{eq:theory-leader-distance}, to smaller values of $\tDL$. Hence, a denser connectivity of the system results in an increasing opinion balance, which can be deduced as follows. While more and more edges are added to the network, the structure gradually turns into a complete graph. By that, the centralities of all potential stubborn agents become increasingly similar, which in turn yields a vanishing mean opinion, cf. Eq.~\eqref{eq:opinion_mean}. Furthermore, $\tilde{D}_\mathrm{max}$ is reduced as the MRD between stubborn agents is decreased due to the addition of links in the network.
The effect of higher levels of stubbornness is illustrated in Fig.~\ref{fig:fig_param_exploration}(c). Except different values for $\kappa$\,, the system's parameters equal those of Fig.~\ref{fig:XYZ}(a). We find that the overall shape of $\tmu$ is approximately retained and its values increase for higher stubbornness. A similar behavior is observed for $\tDL$. For $\kappa\rightarrow\infty$ the averaged maximum opinion spread $\tDL$ approaches the value of $\tDL=2P$ as both stubborn agents get closer to their individual biases. This limit is depicted as the pink dashed line in the bottom panel of Fig.~\ref{fig:fig_param_exploration}(c) and formally derived in \ref{app:DL}.

Finally, we briefly discuss how the structural randomization of an empirical friendship network of $n=70$ nodes and $n_e=274$ edges influences two of the proposed measures to characterize the emerging non-consensus states. Specifically, we investigate how an increasing number of randomly rewired edges $n_r$ affects the mean final opinion $\tilde{\mu}_x$ and the average maximum opinion spread $\tDL$, respectively. The top right panel of Fig.~\ref{fig:fig_empirical_network_quantities} shows the friendship graph, which is characterized by a community structure, resulting from three different school classes \cite{nr_data}. Its original structure gets blurred as $n_r$ is increased, cf. bottom right panels of Fig.~\ref{fig:fig_empirical_network_quantities} for $n_r=200$. 

These structural changes substantially alter the opinion formation processes and the properties of the resulting non-consensus states. We find that an increasing number of rewired edges $n_r$ lowers the value of $\tDL$, cf. left panel of Fig.~\ref{fig:fig_empirical_network_quantities}. Similarly to WS networks the randomization procedure increases the number of shortcuts in the network while local clustering is reduced. This procedure (on average) reduces MRDs between stubborn agents, which in turn leads to a monotonic decrease of $\tDL$.
The absolute mean final opinion $\tmu$ shows a different behavior. After a steep initial increase it reaches a maximum around $n_r=40$. With regard to Eq.~\eqref{eq:opinion_mean} this indicates that the expression $|C(a)-C(b)|/[C(a)C(b)]$, exhibits a non-monotonic behavior as a function of $n_r$.
Note that a similar behavior can be observed in the top panel of Fig.~\ref{fig:fig_param_exploration}(b). For the smallest number of edges ($n_e=150$) $\tmu$ shows a peak at rather low rewiring probabilities. Such behavior might therefore be related to finite-size effects in systems of low average degree, as in the case of the friendship network ($n_e=274$). 

In summary, the systematic evaluation of the expressions derived in Secs.~\ref{subsec:opinion-coherence}-\ref{sec:two-opposing-leaders} shed light on the role of stubborn agents in shaping the opinion formation on general influence networks. The opinion coherence during a consensus change, induced by a single stubborn agent, critically depends on the network structure and the position of the stubborn agent. Furthermore we find that high degree nodes generally favor a coherent transition to the new consensus value. On a network level, opinion coherence is minimal for perfectly symmetric structures (high values of $\tilde{\mathcal{C}}$) while $\tilde{\mathcal{C}}$ decreases upon random rewiring. Furthermore, opinion coherence was decreased (increasing values of $\tilde{\mathcal{C}}$) for increasingly pronounced community structures.  
For a pair of two antagonistic stubborn agents, we have first demonstrated the opinion association problem on different networks. As a simplistic approach to binary social influences acting on a population of agents our formalism is able to capture and explain the opinion associations in terms of MRDs. Beyond that, we have analyzed the properties of emerging non-consensus states. We have found, that the opinion heterogeneity of those states, quantified in terms of $\tilde{D}_\mathrm{max}$ and $\tilde{\sigma}^2_x$, is reduced by a short average path length and increased as the community structure becomes more pronounced. Additionally we have discussed the balancedness of final opinion states in terms of the absolute mean opinion. 
While symmetric ring lattices yield perfectly balanced ($\tilde{\mu}_x=0$) situations, $\tilde{\mu}_x$ generally increases with $p_r$ on WS graphs. On SBM networks $\tilde{\mu}_x$ yields a maximum for intermediate intra-community connection probabilities and rapidly decreases for $p_\mathrm{intra}\rightarrow 1/3$. The size of the system in terms of $n$ and $n_e$ also substantially affects the properties of the emerging opinion states. Focusing on WS networks we find that, the average maximum opinion spread $\tilde{D}_\mathrm{max}$ is decreased if $n$ and $n_e$ are decreased simultaneously. This behavior is reversed, if only the number of edges $n_e$ is reduced. Interestingly, the reduced connectivity leads to larger values of $\tilde{D}_\mathrm{max}$ -- an effect which can be explained by larger MRDs between stubborn agents in the case of low edge densities. Combined with the finding of lower values of $\tilde{\mu}_x$ for high link densities, we conclude that well connected network structures preclude a strong divergence of opinions and increase the balance of opinions in the system.

\section{Conclusions}\label{sec:conclusion}
In this work we consider the Taylor model as an analytically tractable approach to investigate the role of stubborn agents in diffusively coupled populations. Within a Laplacian formalism we treat the dynamics of opinions and the topology of the influence network in a unified framework. We focus on two distinct modes of opinion formation: the change of a prevailing consensus due to a single stubborn agent and the emergence of non-consensus states in the case of two antagonistically biased agents. 

Based on the spectral decomposition of a modified Laplacian of the system, we derive a measure for opinion coherence. It quantifies how coherently a single stubborn agent is able to change a prevailing consensus in the population. In line with previous findings we demonstrate a strong dependence on the position of the stubborn agent. Additionally, we find that a large number of neighbors is beneficial to achieve a high opinion coherence during the transition to a new consensus. In the case of two antagonistically biased agents the system can -- instead -- not reach a global consensus. Hence, we analytically quantify the emerging opinion heterogeneity in terms of the first two moments of the stationary opinion distribution and the maximum opinion spread. In line with previous studies of discrete opinion dynamics \cite{yildiz2011discrete}, we find that the properties of the final states strongly depend on the topology of the influence network and the position of the stubborn agents. Crucially, the developed Laplacian-based formalism allows to formulate these quantities in terms of modified resistance distances, which allow intuitive interpretations of the obtained results on different network topologies. 
The approach is rather general and, hence, not restricted to certain types of influence networks. We note that most of the presented work can be generalized to the case of directed networks. This includes situations in which stubborn agents are not subject to opinion averaging and therefore do not get influenced by the population.

A shortcoming of the presented results is, that the numerical evaluation of the network averaged quantities is computationally expensive. However, certain aspects of our formalism can still be applied to large networks. In particular, setups which do not require an extensive averaging procedure as the number of potentially stubborn agents is small, can be analyzed with reasonable time complexity. Furthermore, the introduced opinion association framework is not restricted to small system sizes as it merely requires a single matrix inversion to determine the MRDs in the system.

Although our work is primarily inspired by social interaction dynamics, the presented formalism might be found useful in various fields dealing with systems of diffusively coupled units.
For example, in electrical networks, the modified Laplacian $\mathbb{L}^{(\kappa)}$ describes the relation between voltages and currents between arbitrary pairs of nodes in the case of networks containing dissipative nodes \cite{dorfler2018electrical}. Hence, the distribution of voltages in such systems can be directly obtained in terms of the modified resistance distances, discussed in this work. 
Furthermore, similar models of distributed consensus formation have been used to study multi-robot formation control~\cite{ren2008distributed} or the collective response in biological swarms~\cite{young2013starling,mateo2019optimal}. Here, the concept of stubborn agents, often called leaders, corresponds to external signals for the purpose of control \cite{Pat18} or might model agents with access to additional information, e.g. cues about potential threats \cite{mateo2019optimal}. Especially, the treatment of such problems in terms of the introduced MRDs might yield insightful and computationally advantageous frameworks to investigate the properties of diffusively coupled systems.

\section{Acknowledgment}
This work was developed within the
scope of the IRTG 1740/TRP 2015/50122-0 and funded
by the DFG/FAPESP and the Swiss National Science Foundation under grant No. 200020\_\!\_182050.
We thank Philipp Lorenz-Spreen and Philippe Jacquod for interesting discussions and Lea Cerekwicki for helpful comments on the manuscript.

\appendix

\section{Proof of equations}
\subsection{}\label{app1}
The product of $\mathbb{L}^{(\kappa)}$ and its inverse reads elementwise,
\begin{eqnarray}
\sum_j \mathbb{L}^{(\kappa)}_{ij} [{\mathbb{L}^{(\kappa)}}]^{-1}_{jk}&=&\sum_j \left(\mathbb{L}_{ij} + \kappa\sum_{a\in V_s}\delta_{ai}\delta_{ij}\right) [{\mathbb{L}^{(\kappa)}}]^{-1}_{jk} \; \\
&=&\delta_{ik} \; ,
\end{eqnarray}
from which, summing over $i$, we have that,
\begin{eqnarray}\label{eq:app1}    
\sum_{a\in V_s}[{\mathbb{L}^{(\kappa)}}]^{-1}_{ja}=1/\kappa \; .
\end{eqnarray}
Both matrices $\mathbb{L}^{(\kappa)}$ 
and its inverse can be written using eigenvectors and eigenvalues as
\begin{eqnarray}\label{eq:eigen}
\mathbb{L}^{(\kappa)}_{ij}&=&\sum_\alpha \lambda_\alpha^{(\kappa)} u_{\alpha,i}^{(\kappa)} u_{\alpha,j}^{(\kappa)} \; ,\\
\left[{\mathbb{L}^{(\kappa)}}\right]_{ij}^{-1}&=&\sum_\alpha \frac{u_{\alpha,i}^{(\kappa)} u_{\alpha,j}^{(\kappa)}}{\lambda_\alpha^{(\kappa)}} \; .
\end{eqnarray}
Thus Eq.~(\ref{eq:1/k}) implies that
\begin{eqnarray}\label{eq:app_k1}
\sum_{a\in V_s}\sum_\alpha\frac{u_{\alpha,k}^{(\kappa)} u_{\alpha,a}^{(\kappa)}}{\lambda_\alpha^{(\kappa)}}=1/\kappa \; .
\end{eqnarray}

\subsection{}\label{app0}
The positivity of the eigenvalues of $\mathbb{L}^{(\kappa)}$ follows from
\begin{align}
\begin{split}
{\bf u}_{\alpha}^{(\kappa)}{}^T \mathbb{L}^{(\kappa)} {\bf u}_{\alpha}^{(\kappa)} &=\sum_{i,j}u^{(\kappa)}_{\alpha,i} \left(\mathbb{L}_{ij} + \kappa\sum_{a\in V_s}\delta_{ai}\delta_{ij}\right) u^{(\kappa)}_{\alpha,i}\\
&= {\bf u}^{(\kappa)}_{\alpha} \mathbb{L} {\bf u}^{(\kappa)}_{\alpha} + \kappa\sum_{a\in V_s} {u^{(\kappa)}_{\alpha,a}}^2  \; ,\\
&= \lambda_\alpha^{(\kappa)} > 0 \; ,
\end{split}
\end{align}
where in the last line we used that the Laplacian $\mathbb{L}$ is positive semidefinite.

\subsection{}\label{app2}
In the case of two stubborn agents $V_s=\{a,b\}$, we have for the element $(a,b)$ of the product of $\mathbb{L}^{(\kappa)}$ by its inverse,
\begin{eqnarray}
\sum_j \mathbb{L}_{aj}^{(\kappa)}[{\mathbb{L}^{(\kappa)}}]_{jb}^{-1}&=&\sum_j
\mathbb{L}_{aj}[{\mathbb{L}^{(\kappa)}}]_{jb}^{-1}+ \kappa [{\mathbb{L}^{(\kappa)}}]_{ab}^{-1}=0 \;.
\end{eqnarray}
Exploiting the symmetry of matrices ${\mathbb{L}^{(\kappa)}}^{-1}$ and $\mathbb{L}$ together with Eq.~(\ref{eq:app1}) we get,
\begin{eqnarray}\label{eq:prop22}
[{\mathbb{L}^{(\kappa)}}]_{aa}^{-1}&=&[{\mathbb{L}^{(\kappa)}}]_{bb}^{-1}\; ,\\
\sum_j \mathbb{L}_{aj}[{\mathbb{L}^{(\kappa)}}]_{jb}^{-1}&=&\sum_j \mathbb{L}_{bj}[{\mathbb{L}^{(\kappa)}}]_{ja}^{-1}\; .
\end{eqnarray}
With Eq.~(\ref{eq:eigen}) we have the relation between eigenvectors and eigenvalues,
\begin{eqnarray}
\sum_\alpha\frac{{u_{\alpha,a}^{(\kappa)}}^2}{\lambda_\alpha^{(\kappa)}}&=&\sum_\alpha\frac{{u_{\alpha,b}^{(\kappa)}}^2}{\lambda_\alpha^{(\kappa)}} \; .
\end{eqnarray}

\subsection{}\label{App4}
Multiplying $\mathbb{L}^{(\kappa)}$ by eigenvector ${\bf u}_{\alpha}^{(\kappa)}$ reads elementwise
\begin{eqnarray}
\sum_j\mathbb{L}^{(\kappa)}_{ij}u_{\alpha,j}^{(\kappa)}&=&\sum_j \left(\mathbb{L}_{ij}+\delta_{ij}\kappa\sum_{a\in V_s}\delta_{aj}\right)u_{\alpha,j}^{(\kappa)}\\
&=&\lambda_\alpha^{(\kappa)} u_{\alpha,i}^{(\kappa)} \; .
\end{eqnarray}
Summing over $i$ we get
\begin{eqnarray}
\kappa\sum_j \sum_{a\in V_s}\delta_{aj}\, u_{\alpha,j}^{(\kappa)}=\lambda_\alpha^{(\kappa)} \sum_i u_{\alpha,i}^{(\kappa)} \; 
\end{eqnarray}
which finally yields
\begin{eqnarray}\label{eq:app_k2}
\kappa\sum_{a\in V_s} \frac{u_{\alpha,a}^{(\kappa)}}{{\lambda_\alpha^{(\kappa)}}}= \sum_i u_{\alpha,i}^{(\kappa)} \; .
\end{eqnarray}

\subsection{}\label{app:leada}
The final state of opinions $\mathbf{x}(t\rightarrow\infty)$ are obtained by a spectral decomposition over the eigenvectors of $\mathbb{L}^{(\kappa)}$, cf. Eq.~\eqref{eq:spec-decomp}. In the case of two antagonistic stubborn agents $a$, $b$ such that $P_a=-P_b=P$, the expansion coefficients, Eq.~(\ref{eq:sol}), with $\kappa P_i= \kappa P(\delta_{ia}-\delta_{ib})$ and vanishing initial conditions (full consensus) are given by
\begin{align}
\begin{split}
c_\alpha(t\rightarrow \infty) = \kappa P\frac{(u_{\alpha,a}^{(\kappa)}-u_{\alpha,b}^{(\kappa)})}{\lambda_\alpha^{(\kappa)}} \;,\quad \alpha=1,...,n\, .
\end{split}
\end{align}
The final opinion of agent $i$ thus reads,
\begin{eqnarray}
x_i^\infty &=& \sum_\alpha c_\alpha(t\rightarrow\infty)u_{\alpha,i}^{(\kappa)}\\
&=& \kappa P\sum_\alpha\frac{(u_{\alpha,a}^{(\kappa)}-u_{\alpha,b}^{(\kappa)})}{\lambda_\alpha^{(\kappa)}}u_{\alpha,i}^{(\kappa)} \; .
\end{eqnarray}
One can express the latter equation with MRD Eq.~(\ref{eq:RLd}) as follows:
\begin{eqnarray}
x_i^\infty &=& \frac{\kappa P}{2}\sum_\alpha \frac{(2u_{\alpha,a}^{(\kappa)} u_{\alpha,i}^{(\kappa)}-2u_{\alpha,b}^{(\kappa)} u_{\alpha,i}^{(\kappa)})}{\lambda_\alpha^{(\kappa)}}\\
&=&  \frac{\kappa P}{2}[\Omega^{(\kappa,1)}_{bi}(\{a,b\})-\Omega^{(\kappa,1)}_{ai}(\{a,b\})]\; ,
\end{eqnarray}
where we used Eq.~\eqref{eq:prop2}, namely that 
$\sum_\alpha \frac{{u_{\alpha,a}^{(\kappa)}}^2}{\lambda_\alpha^{(\kappa)}}=\sum_\alpha \frac{{u_{\alpha,b}^{(\kappa)}}^2}{\lambda_\alpha^{(\kappa)}}$.

\subsection{}\label{app:DL}
In the case of two antagonistic stubborn agents $a$, $b$ with opinions $P_a=-P_b=P$, the final opinion of agent $i$ is given by Eq.~(\ref{eq:final-state-x_i-two-leaders}) and reads:
\begin{eqnarray}\label{eq:appfinal2}
x_i^{\infty}=\sum_{\alpha}\frac{u_{\alpha,a}^{(\kappa)}-u_{\alpha,b}^{(\kappa)}}{\lambda_\alpha^{(\kappa)}}\, u_{\alpha,i}^{(\kappa)} \; .
\end{eqnarray}
The distance between the two stubborn agents in the final state is given by,
\begin{eqnarray}
D_\mathrm{max} &=& x_a^\infty-x_b^\infty  \\ 
&=&\kappa P \sum_{\alpha}\frac{u_{\alpha,a}^{(\kappa)}-u_{\alpha,b}^{(\kappa)}}{\lambda_\alpha^{(\kappa)}}(u_{\alpha,a}^{(\kappa)}-u_{\alpha,b}^{(\kappa)}) \; \\
&=&\kappa P\Omega_{ab}^{(\kappa,1)}\; ,
\end{eqnarray}
where we used the definition of MRD, Eq.~(\ref{eq:eig}).

\subsection{}\label{app:stub}
In the case of two antagonistic stubborn agents with finite stubbornness $\kappa$, the opinion distance between the pair of stubborn agents satisfies $D_\mathrm{max}\le|P_a-P_b|$. For $\kappa\rightarrow\infty$, each stubborn agent reaches his/her own final bias and thus the distance between them should satisfy $D_\mathrm{max}=|P_a-P_b|$. The latter can be shown as follows. From Eqs.~(\ref{eq:app_k1}), (\ref{eq:app_k2}) one has
\begin{eqnarray}
\sum_\alpha\frac{{u_{\alpha,a}^{(\kappa)}}^2+u_{\alpha,a}^{(\kappa)} u_{\alpha,b}^{(\kappa)}}{\lambda_\alpha^{(\kappa)}}&=&1/\kappa \; ,\\
\sum_\alpha\frac{{u_{\alpha,b}^{(\kappa)}}^2+u_{\alpha,a}^{(\kappa)} u_{\alpha,b}^{(\kappa)}}{\lambda_\alpha^{(\kappa)}}&=&1/\kappa \; ,\\
\frac{{u_{\alpha,a}^{(\kappa)}} + {u_{\alpha,b}^{(\kappa)}}}{\lambda_\alpha^{(\kappa)}}&=&\kappa^{-1}\sum_i{u_{\alpha,i}^{(\kappa)}} \; .
\end{eqnarray}
Then using the definition of MRD Eq.~(\ref{eq:eig}) with $p=1$ one has,
\begin{eqnarray}
 D_\mathrm{max} &=& \kappa P \Omega_{ab}^{(\kappa,1)}\\
&=& \kappa P \sum_\alpha \frac{(u_{\alpha,a}^{(\kappa)} -u_{\alpha,b}^{(\kappa)})^2}{\lambda_\alpha^{(\kappa)}} \\
&=&\kappa P \sum_\alpha \frac{{u_{\alpha,a}^{(\kappa)}}^2 -2{u_{\alpha,a}^{(\kappa)}}{u_{\alpha,b}^{(\kappa)}} + {u_{\alpha,b}^{(\kappa)}}^2}{\lambda_\alpha^{(\kappa)}} \\
&=&\kappa P \left[ 2\sum_\alpha \frac{{u_{\alpha,a}^{(\kappa)}}^2+{u_{\alpha,b}^{(\kappa)}}^2}{\lambda_\alpha^{(\kappa)}} -\frac{2}{\kappa} \right] \; . 
\end{eqnarray}
Finally, taking stubbornness $\kappa$ to infinity one has
\begin{eqnarray}
\lim_{\kappa\rightarrow \infty} D_\mathrm{max} &=& 2P \; ,
\end{eqnarray}
where we used $\lim_{\kappa\rightarrow \infty}\sum_\alpha \frac{{u_{\alpha,a}^{(\kappa)}}^2}{\lambda_\alpha^{(\kappa)}}=1/\kappa$\,.

\subsection{}\label{app:mu_sigma}
To obtain the mean value of the opinions in the final state, one has to average Eq.~(\ref{eq:appfinal2}) over all agents yielding
\begin{align}
\begin{split}\label{eq:final-state-x_i-two-leaders_app}
\mu_x(\{a,b\})&=n^{-1}\sum_i x^{\infty}_i\\
&=n^{-1}\sum_i\frac{\kappa P}{2}[\Omega^{(\kappa,1)}_{bi}(\{a,b\})-\Omega^{(\kappa,1)}_{ai}(\{a,b\})]\\
&= \frac{\kappa P}{2}[C_1^{-1}(b)-C_1^{-1}(a)]\;,
\end{split}
\end{align}
where in the last equality we used the definition of centralities Eq.~(\ref{eq:centr}). The variance of the opinions in the final state are given by
\begin{align}
\begin{split}
\sigma_x^2(\{a,b\}) &= n^{-1}\sum_i (x_i^\infty-\mu_x)^2\\
&= \left( \frac{\kappa P}{2} \right)^2\sum_i \left[ \Omega_{ib}^{(\kappa,1)} -\Omega_{ia}^{(\kappa,1)}\right.\\
&\left.- C_1^{-1}(b) + C_1^{-1}(a) \right]^2 \; . 
\end{split}
\end{align}
Then, to obtain Eq.~(\ref{eq:opinion_variance}), we note that
\begin{align}
\frac{1}{4}\sum_i \left( \Omega_{ib}^{(\kappa,1)} - \Omega_{ia}^{(\kappa,1)} \right)^2 &= \sum_{i,\alpha} \frac{(-u_{\alpha,i}^{(\kappa)} u_{\alpha,b}^{(\kappa)} + u_{\alpha,i}^{(\kappa)} u_{\alpha,a}^{(\kappa)})^2}{{\lambda_\alpha^{(\kappa)}}^2} \\
&= \Omega_{ab}^{(\kappa,2)} \; .
\end{align}
In the last equality we used Eq.~(\ref{eq:prop2}), namely $\sum_\alpha \frac{{u_{\alpha,a}^{(\kappa)}}^2}{\lambda_\alpha^{(\kappa)}}=\sum_\alpha \frac{{u_{\alpha,b}^{(\kappa)}}^2}{\lambda_\alpha^{(\kappa)}}$.

\section{}\label{app:network-models-dataset}
\textbf{Watts-Strogatz (WS) model.} Introduced in \cite{watts1998collective}, the WS
model constructs networks as follows.
Initially a ring of $n$ nodes, where each 
node is connected to its $K_{WS}$ nearest neighbors, is established.
Subsequently, each edge is rewired with probability $p_r$.
Typically, $p_r$ interpolates between 
a regular ring lattice and a random network.

\textbf{Stochastic block model (SBM):} The basic concept of the SBM was originally introduced in the social sciences \cite{holland1983stochastic}. The principle is that nodes of a network are organized into groups. Subsequently, connection probabilities for pairs of nodes within and between those groups are defined. For the simple case considered in this work we assume two groups of equal size ($n/2$) and define $p_\mathrm{intra}$ and $p_\mathrm{inter}$, as the probabilities for links within and across groups, respectively.
In order to tune the community structure of the network, while fixing the average degree, we implement the following relation 
\begin{equation}\label{eq:app_sbm}
    p_\mathrm{inter} = \frac{n_e-[(n/2)^2-n/2]\,p_\mathrm{intra}}{(n/2)^2}\,.
\end{equation}
Note that in the limit of $p_\mathrm{intra}\rightarrow 1/3$ the SBM network disintegrated into two separate subgraphs as $p_\mathrm{inter}\rightarrow 0$ in the considered case of $n=50$ and $n_e=200$.

\textbf{Barabasi-Albert (BA) model.} 
Networks generated according to the BA model are constructed using the preferential attachment rule \cite{barabasi1999emergence}. In our implementation a network of $n$ nodes is grown by attaching a newly introduced node with $m$ edges. The preferential attachment mechanism ensures that the probability for a new node establishing a connection with an existing one is proportional to the degree of
the latter.

\textbf{Empirical friendship network.} The data set contains information about friendship relations within a US highschool \cite{nr_data}. 
To construct a friendship network the students were asked twice about their friends. The original network is directed and weighted to account for multiple namings of a single student by a friend. For our purposes we symmetrize the interaction topology $\mathbf{B}$ and dismiss weights such that we have $b_{ij}=1$ if one of the two students ($i,j$) named the other as a friend and $b_{ij}=0$, otherwise. The analyzed network contains $n=70$ nodes and $n_e=274$ edges. To randomize the network topology we perform an increasing number of double edge swaps. This procedure fixes the degrees of the nodes but randomizes the connectivity structure \cite{fosdick2018configuring}. 

\section{}\label{app:App_merge}
\begin{figure}[h!]
\includegraphics[width=.99\linewidth]{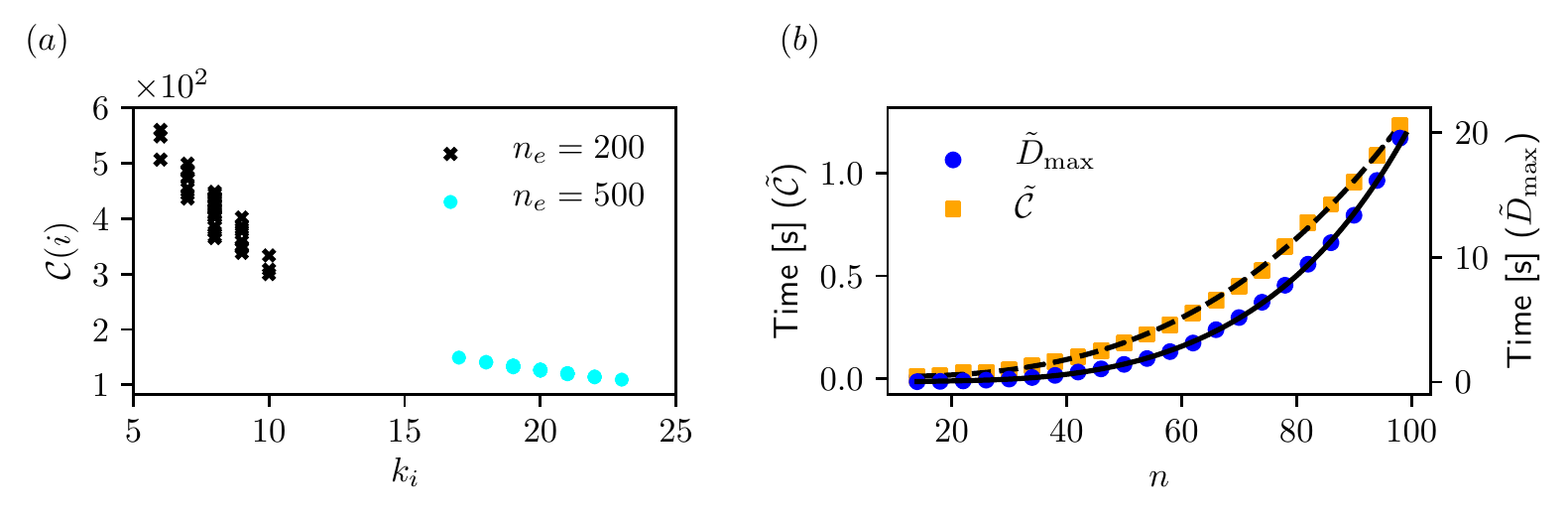}
\caption{Panel (a): relation between the coherence measure $\mathcal{C}(i)$ and the degree $k_i$ of node $i$ on which the single stubborn agent is placed. The depicted results correspond to two WS networks with $n=50$ and a different number of edges $n_e$. The rewiring probability is $p_r=0.2$ in both cases. Panel (b): time complexity in seconds $[s]$ for the computation of the quantities $\tC$ (orange squares) and $\tDL$ (blue dots) as a function of the network size $n$. The black dashed and solid lines correspond to polynomial fits of the orders $\mathcal{O}(n^3)$ ($\tilde{\mathcal{C}}$) and $\mathcal{O}(n^5)$ ($\tilde{D}_\mathrm{max}$), respectively.}
\label{fig:fig8_App_physA}
\end{figure}

In Fig.~\ref{fig:fig8_App_physA}(a) we depict the relation between the opinion coherence $\mathcal{C}(i)$ and the degree $k_i$ for a stubborn agent placed on node $i$ on two different WS networks. The black crossed markers correspond to the results depicted in the top panel of Fig.~\ref{fig:fig_coherence_map}(b). The cyan dots show results of a WS network with an increased number of $n_e=500$ edges.

In Fig.~\ref{fig:fig8_App_physA}(b) we show the time complexity for the numerical evaluation of $\tC$ and $\tDL$. The quantities are computed as averages over all possible sets of stubborn agents on WS networks with $n$ nodes and $K_\mathrm{WS}=4$. Due to the quadratic growth of the possible sets $V_s$ as a function of $n$ for $n_s=2$ the time complexity for computing $\tDL$ (blue dots) is increased about a factor of $n^2$ compared to $\tC$ (orange squares).

\end{document}